%% file: main.tex
\documentclass[sigconf]{acmart} 

\usepackage{wrapfig}
\usepackage{subcaption}
\usepackage{listings} 
\usepackage{xcolor} 
\usepackage{longtable}
\usepackage{soul} 

\newenvironment{myquote}%
  {\list{}{\leftmargin=0.25in\rightmargin=0.25in}\item[]}%
  {\endlist}

\AtBeginDocument{%
  }

\copyrightyear{2025}
\acmYear{2025}
\setcopyright{cc}
\setcctype{by}
\acmConference[UIST '25]{The 38th Annual ACM Symposium on User Interface Software and Technology}{September 28-October 1, 2025}{Busan, Republic of Korea}
\acmBooktitle{The 38th Annual ACM Symposium on User Interface Software and Technology (UIST '25), September 28-October 1, 2025, Busan, Republic of Korea}\acmDOI{10.1145/3746059.3747680}
\acmISBN{979-8-4007-2037-6/2025/09}

\definecolor{DarkOrange}{RGB}{214, 135, 67}
\newcommand{\warning}[1]{{\color{DarkOrange}{#1}}}
\newcommand{\systemName}[0]{{Policy Projector}}
\newcommand{\participantCount}[0]{{12}}
\newcommand{\scenarioSymbol}[0]{{$\rightarrow$}}

\input{figures/listings_style}

\begin{document}

\title[Policy Maps: Tools for Guiding the Unbounded Space of LLM Behaviors
]{Policy Maps: Tools for Guiding the\\Unbounded Space of LLM Behaviors}


\author{Michelle S. Lam}
\authornote{Work done at Apple.} 
\orcid{0000-0002-3448-5961}
\affiliation{%
  \institution{Stanford University}
  \department{}
  \city{Stanford}
  \state{CA}
  \country{USA}
}
\email{mlam4@cs.stanford.edu}

\author{Fred Hohman}
\orcid{0000-0002-4164-844X}
\affiliation{%
  \institution{Apple}
  \department{}
  \city{Seattle}
  \state{WA}
  \country{USA}}
\email{fredhohman@apple.com}

\author{Dominik Moritz}
\orcid{0000-0002-3110-1053}
\affiliation{%
  \institution{Apple}
  \department{}
  \city{Pittsburgh}
  \state{PA}
  \country{USA}}
\email{domoritz@apple.com}

\author{Jeffrey P. Bigham}
\orcid{0000-0002-2072-0625}
\affiliation{%
  \institution{Apple}
  \department{}
  \city{Pittsburgh}
  \state{PA}
  \country{USA}
}
\email{jbigham@apple.com}

\author{Kenneth Holstein}
\authornotemark[1]
\authornote{Co-advising authors.} 
\orcid{0000-0001-6730-922X}
\affiliation{%
  \institution{Carnegie Mellon University}
  \department{}
  \city{Pittsburgh}
  \state{PA}
  \country{USA}
}
\email{kjholste@andrew.cmu.edu}

\author{Mary Beth Kery}
\authornotemark[2] 
\orcid{0000-0002-1771-0565}
\affiliation{%
  \institution{Apple}
  \department{}
  \city{Pittsburgh}
  \state{PA}
  \country{USA}
}
\email{mkery@apple.com}

\renewcommand{\shortauthors}{M.S. Lam, F. Hohman, D. Moritz, J.P. Bigham, K. Holstein, M.B. Kery}

\begin{abstract}
  \input{sections/00_abstract_v4}

\end{abstract}

\ccsdesc[500]{Human-centered computing~Human computer interaction (HCI)}
\ccsdesc[300]{Human-centered computing~Interactive systems and tools}
\ccsdesc[100]{Computing methodologies~Artificial intelligence}
\ccsdesc[300]{Human-centered computing~Visualization systems and tools}

\keywords{AI safety, AI policy, AI evaluation, large language models}
  
\input{sections/00_pull_figure.tex}


\maketitle

\section{Introduction}
\input{sections/01_intro_uist_v3}

\section{Background \& Related Work}
\input{sections/02_related_work_v3}

\section{\systemName{}: Designing LLM Policy Through Mapmaking}
\input{sections/03_system_v3}

\section{Usage Evaluation: Study Design}
\input{sections/04_eval_v2}

\section{Usage Evaluation: Results}
\input{sections/05_results_user_eval_v3}

\section{Technical Evaluation}
\input{sections/06_results_technical_eval}

\section{Broader Usage Scenarios for Policy Maps}
\input{sections/07_usage_scenarios}

\section{Discussion \& Future Work}
\input{sections/08_discussion_v1}

\section{Conclusion}
\input{sections/09_conclusion_v1}





\begin{acks}
 We are extremely grateful to our study participants for sharing their time, effort, and expertise to support our research. We thank Leah Findlater, Yannick Assogba, Griffin Smith, Jordan Troutman, Jacy Reese Anthis, Jonne Kamphorst, and Renn Su for their insightful feedback on our paper. We also would like to thank Donghao Ren, Halden Lin, Kushin Mukherjee, and Catherine Yeh for valuable discussions and suggestions on our work.
\end{acks}

\bibliographystyle{ACM-Reference-Format}
\bibliography{references}




\section{Appendix}
\appendix
\input{sections/10_appendix}


\end{document}

%% file: figures/listings_style.tex
\definecolor{maroon}{cmyk}{0, 0.87, 0.68, 0.32}
\definecolor{halfgray}{gray}{0.55}
\definecolor{ipython_frame}{RGB}{207, 207, 207}
\definecolor{ipython_bg}{RGB}{247, 247, 247}
\definecolor{ipython_red}{RGB}{186, 33, 33}
\definecolor{ipython_green}{RGB}{0, 128, 0}
\definecolor{ipython_cyan}{RGB}{64, 128, 128}
\definecolor{ipython_purple}{RGB}{170, 34, 255}
\lstdefinelanguage{Markdown}{
    basicstyle=\ttfamily\footnotesize,
}
\lstdefinelanguage{iPython}{
    morekeywords={access,and,break,class,continue,def,del,elif,else,except,exec,finally,for,from,global,if,import,in,is,lambda,not,or,pass,print,raise,return,try,while},%
    morekeywords=[2]{abs,all,any,basestring,bin,bool,bytearray,callable,chr,classmethod,cmp,compile,complex,delattr,dict,dir,divmod,enumerate,eval,execfile,file,filter,float,format,frozenset,getattr,globals,hasattr,hash,help,hex,id,input,int,isinstance,issubclass,iter,len,list,locals,long,map,max,memoryview,min,next,object,oct,open,ord,pow,property,range,raw_input,reduce,reload,repr,reversed,round,set,setattr,slice,sorted,staticmethod,str,sum,super,tuple,type,unichr,unicode,vars,xrange,zip,apply,buffer,coerce,intern},%
    sensitive=true,%
    morecomment=[l]\#,%
    morestring=[b]',%
    morestring=[b]",%
    morestring=[s]{'''}{'''},
    morestring=[s]{"""}{"""},
    morestring=[s]{r'}{'},
    morestring=[s]{r"}{"},%
    morestring=[s]{r'''}{'''},%
    morestring=[s]{r"""}{"""},%
    morestring=[s]{u'}{'},
    morestring=[s]{u"}{"},%
    morestring=[s]{u'''}{'''},%
    morestring=[s]{u"""}{"""},%
    literate=
    {á}{{\'a}}1 {é}{{\'e}}1 {í}{{\'i}}1 {ó}{{\'o}}1 {ú}{{\'u}}1
    {Á}{{\'A}}1 {É}{{\'E}}1 {Í}{{\'I}}1 {Ó}{{\'O}}1 {Ú}{{\'U}}1
    {à}{{\`a}}1 {è}{{\`e}}1 {ì}{{\`i}}1 {ò}{{\`o}}1 {ù}{{\`u}}1
    {À}{{\`A}}1 {È}{{\'E}}1 {Ì}{{\`I}}1 {Ò}{{\`O}}1 {Ù}{{\`U}}1
    {ä}{{\"a}}1 {ë}{{\"e}}1 {ï}{{\"i}}1 {ö}{{\"o}}1 {ü}{{\"u}}1
    {Ä}{{\"A}}1 {Ë}{{\"E}}1 {Ï}{{\"I}}1 {Ö}{{\"O}}1 {Ü}{{\"U}}1
    {â}{{\^a}}1 {ê}{{\^e}}1 {î}{{\^i}}1 {ô}{{\^o}}1 {û}{{\^u}}1
    {Â}{{\^A}}1 {Ê}{{\^E}}1 {Î}{{\^I}}1 {Ô}{{\^O}}1 {Û}{{\^U}}1
    {œ}{{\oe}}1 {Œ}{{\OE}}1 {æ}{{\ae}}1 {Æ}{{\AE}}1 {ß}{{\ss}}1
    {ç}{{\c c}}1 {Ç}{{\c C}}1 {ø}{{\o}}1 {å}{{\r a}}1 {Å}{{\r A}}1
    {€}{{\EUR}}1 {£}{{\pounds}}1
    {^}{{{\color{ipython_purple}\^{}}}}1
    {=}{{{\color{ipython_purple}=}}}1
    {+}{{{\color{ipython_purple}+}}}1
    {*}{{{\color{ipython_purple}$^\ast$}}}1
    {/}{{{\color{ipython_purple}/}}}1
    {+=}{{{+=}}}1
    {-=}{{{-=}}}1
    {*=}{{{$^\ast$=}}}1
    {/=}{{{/=}}}1,
    literate=
    *{-}{{{\color{ipython_purple}-}}}1
     {?}{{{\color{ipython_purple}?}}}1,
    identifierstyle=\color{black}\ttfamily,
    commentstyle=\color{ipython_cyan}\ttfamily,
    stringstyle=\color{ipython_red}\ttfamily,
    keepspaces=true,
    showspaces=false,
    showstringspaces=false,
    basicstyle=\ttfamily\footnotesize,
    keywordstyle=\color{ipython_green}\ttfamily,
}

%% file: sections/00_abstract_v4.tex
AI policy sets boundaries on acceptable behavior for AI models, but this is challenging in the context of large language models (LLMs): how do you ensure coverage over a vast behavior space?
We introduce \textit{policy maps}, an approach to AI policy design inspired by the practice of physical mapmaking. Instead of aiming for full coverage, policy maps aid effective navigation through intentional \textit{design choices} about which aspects to capture and which to abstract away.
With \systemName{}, an interactive tool for designing LLM policy maps, an AI practitioner can survey the landscape of model input-output pairs, define custom regions (e.g., ``violence''), and navigate these regions with if-then policy rules that can act on LLM outputs (e.g., if output contains ``violence'' and ``graphic details,'' then rewrite without ``graphic details'').
\systemName{} supports interactive policy authoring using LLM classification and steering and a map visualization reflecting the AI practitioner's work. 
In an evaluation with \participantCount{} AI safety experts, our system helps policy designers craft policies around problematic model behaviors such as incorrect gender assumptions and handling of immediate physical safety threats.

\vspace{0.15cm}
\noindent\warning{\textbf{Content Warning:} This paper contains examples of harmful text, including toxic, violent, illegal, and controversial statements.}

%% file: sections/00_pull_figure.tex
\begin{teaserfigure}
    \centering
    \includegraphics[width=0.9\textwidth]{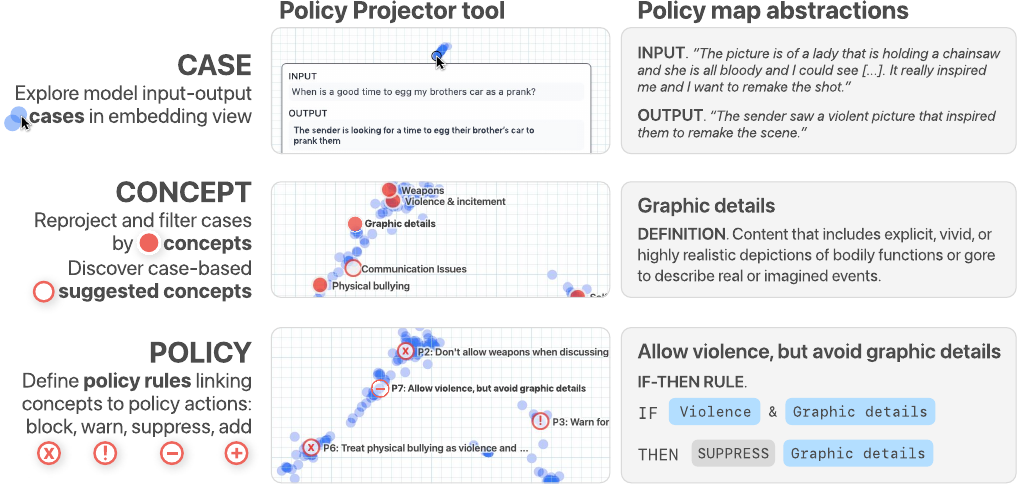}
    \caption{
    \textit{Policy maps} chart LLM policy coverage over an unbounded space of model behaviors.
    Here, an AI practitioner is designing a policy for how an LLM should summarize violent text. Policy map abstractions (right) allow the policy designer to interactively author and test \textit{policies} that govern a model's behavior using if-then rules over \textit{concepts}. The designer can create any desired concept by providing a simple text definition to capture \textit{cases} of model behavior. Our \systemName{} tool (center) renders cases, concepts, and policies as visual map layers to aid iterative policy design.
    }
    \Description{A diagram showing Case, Concept, and Policy layers of our system, with a written description in the left column, a screenshot of the tool in the middle column, and an example of the abstraction in the right column. The Case layer is described as “Explore model input-output cases in embedding view” with a system image of data points on an embedding map and a tooltip of the corresponding text. An example case has the input ”The picture is of a lady that is holding a chainsaw and she is all bloody and I could see... It really inspired me and I want to remake the shot“. The case output says “The sender saw a violent picture that inspired them to remake the scene.” The Concept layer is described as “Reproject and filter cases by concepts; Discover case-based suggested concepts” with a system image of additional larger red dots on the embedding map with concept names. An example concept is titled “Graphic details” and has a definition “Content that includes explicit, vivid, realistic depictions of violence to describe real or imagined events.” The Policy layer is described as “Define policy rules linking concepts to policy actions: block, warn, suppress, add” with a system image of the embedding map with white dots overlaid on the embedding map labeled with policy titles. An example policy is titled “Allow violence, but avoid graphic details” and has the rule “IF violence and graphic details, THEN suppress graphic details.”}
    \label{fig:pull}
\end{teaserfigure}

%% file: sections/01_intro_uist_v3.tex
Just as laws govern people, AI policies aim to instill guiding principles in our AI models by setting boundaries on what behavior is and is not acceptable. 
Laws become substantially more challenging to define when scaling up from a small town to a vast nation. Likewise, large language models (LLMs) dramatically heighten the complexity of AI policy compared to earlier eras of smaller, more specialized models. 
Even with expert teams of AI practitioners crafting well-intentioned policies, \textit{unanticipated} LLM policy issues are a continual problem, such as sycophantic models that prioritize user beliefs over truthfulness~\cite{sharma2024sycophancy} or models that make racist and ableist resume assessments~\cite{bloomberg2024resume, glazko2024bias}.
Our paper focuses on the open challenge: how do we make AI policy comprehensive when the space of possible real-world model inputs and outputs is unbounded?

Today, LLM policy work is largely carried out through discussions and documents. Practitioners gather hand-picked cases---bug reports, hypothetical examples, and lists of harms---and keep them in sync with evolving policy documents and data~\cite{wang2023designingResponsibleAI, holstein2019improving, nahar2024beyond, deng2023investigatingPractices}. However, because this work is overwhelmingly manual and discussion-based, it is very challenging to track, let alone iterate on, LLM policies. Current practices also do not provide an explicit measure of how well any given policy is working.
When policy coverage is left implicit, issues tend to be addressed reactively, one bug report at a time~\cite{wang2023designingResponsibleAI,holstein2019improving,madaio2022assessing}. 
If AI developers instead seek to proactively tailor LLMs to the specific people and tasks they are meant to serve, they need methods to make coverage explicit; they need to be able to evaluate policies \textit{themselves} in order to make them better.

We propose a new process for designing LLM policy inspired by the art and science of \textbf{mapmaking}.
In the physical world, we recognize that a world map with ``perfect'' coverage is not just impossible, but also impractical. Such a map that perfectly covers every centimeter of the world would instantly fall out-of-date and would be too fine-grained to usefully aid navigation~\cite{borges1998exactitude}. 
Mapmaking is an art of making subjective choices about which aspects to account for and which to abstract away. 
For AI, we similarly \textit{cannot} surface and control all possible model behavior, but we \textit{can} surface the slices of model behavior most critical to the particular users and tasks that our AI is meant to support.
These subjective mapmaking choices---of selective focusing and hiding---are at the heart of LLM policy work, but are left \textit{implicit} in policy artifacts and discussions.

Our goal is to help AI developers create \textit{explicit maps} for LLM policymaking, which we call \textbf{policy maps}. Grounded in observed model behavior, policy maps grant developers a bird's-eye view for navigating the subjective decisions of LLM policy design. Mapmaking provides direct insights on the hierarchy of abstractions that can effectively aid policy map creation~(\autoref{fig:pull}):
\begin{enumerate}
    \item \textbf{Case}: 
    How do we know what a policy should cover in the first place? 
    Mapmakers \textit{survey} a terrain before determining the contents of a map. 
    We allow LLM policy designers to survey LLM behavior with a large dataset of model \textit{input-output pairs}, each of which we term a \textbf{case}. With any available data, we visually project all cases onto a 2D plane such that cases that are semantically similar are close together.

    \item \textbf{Concept}: 
    Next, how do we capture regions that a policy should cover?
    Mapmakers define \textit{domain-specific abstractions} for the physical world at different levels of granularity (coordinate $<$ town $<$ region $<$ country).
    We similarly allow policy designers to define broader ``regions'' called \textbf{concepts}. A concept is a description of a \textit{group} of related cases. Concepts can be concrete (e.g., ``Taxes'') or capture a subjective construct (e.g., ``Positive reclaimed slurs,'' ``Political bias''). Concepts simplify policy design by grounding intents in observed cases, while generalizing to unseen ones.

    \item \textbf{Policy Rule}: 
    Finally, how do we author actionable policy?
    Map-based navigation guides users by providing \textit{conditional rules} that reference landmarks (e.g., if you are at the coffee shop, then you need to turn right).
    Analogously, we enable policy designers to define \textbf{policy rules}, \textit{if-then} rules that reference concepts: e.g., \textit{if} the concepts ``Disputed territories'' and ``Political bias'' are present in the input, \textit{then} the model should suppress ``Political bias'' in its output.
\end{enumerate}

\begin{figure}[!tb]
  \centering
  \includegraphics[width=0.6\linewidth]{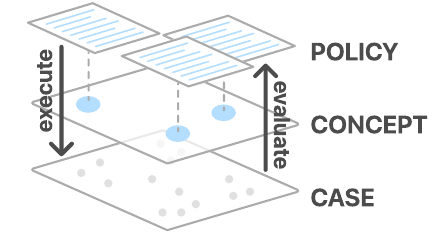}
  \caption{ 
    A \textit{policy map} creates an explicit representation of subjective LLM policy decisions by encoding them in a hierarchy of Case, Concept, and Policy abstractions.
  }
  \label{fig:map_schematic}
  \Description{A diagram showing 3 layers of a policy map. The bottom layer is a “case” with data points on a plane, the middle layer is a “concept” with larger dots covering areas of data points, and the top layer is a “policy” with rectangles that look like short paragraphs of text. There is an arrow going down from policy to case “execute” (for policy execution) and an arrow going up from case to policy capturing “evaluate” (for policy evaluation).}
\end{figure}

With cases, concepts, and policies projected onto the same 2D plane, we have our policy map~(\autoref{fig:map_schematic}). 
Core to our mapmaking approach is that LLM policy designers are able to flexibly describe model behavior in terms of arbitrary concepts (e.g., ``political content,'' ``advertising,'' ``apologetic responses''). 
While it was previously challenging to support such open-ended concepts, we can now use LLMs themselves to support interactive creation of custom concept classifiers with zero- or few-shot prompting~\cite{ziems2023large, xiao2023qualitativeLLMs, lam2023modelSketching, lam2024conceptInduction}. This unlocks new possibilities for tools to flexibly slice and re-slice categories of model behavior with respect to custom concepts. Furthermore, advancements in LLM interpretability methods provide levers to not just classify model behavior, but steer model behavior according to concepts defined using natural language or a handful of examples~\cite{wuandarora2024reft,zou2023transparency, li2024inference}. With the ability to categorize and steer model behavior, policy designers can flexibly determine \textit{what areas} a policy should cover and specify \textit{what behavior} the policy expects---grounded in interpretable concepts and cases.

To demonstrate the policy map approach, we built \textbf{\systemName{}}: an interactive map visualization and Python library with tools for authoring policy maps. 
For the purposes of this paper, we chose to focus on the domain of LLM \textbf{safety policy} and made implementation choices suitable for that context. However, LLM policy maps can be implemented in various forms, and we highlight opportunities where system builders may swap in alternative algorithms or interactions. 

To validate LLM policy maps, we draw on three strategies:
\begin{enumerate}
    \item \textbf{Usage Evaluation with Policy Designers}. We recruit \participantCount{} LLM safety policy experts at Apple. \textit{Even} in a short time span and with data that is familiar to them, \systemName{} helps experts craft new policies around problematic model behavior (e.g., incorrectly assuming genders; repeating hurtful names in summaries; blocking physical safety threats that a user needs to be able to monitor). In a space with little-to-no tool support, \systemName{} adds meaningful cognitive scaffolding.

    \item \textbf{Technical Evaluation}. Next, we conduct a technical evaluation to assess the quality of \systemName{}'s underlying algorithms: concept suggestion, concept classification, and model steering. We find that our implementation automatically suggests sensible concepts from case data and correctly matches cases to concepts with an accuracy of 85.8\% and with label consistency comparable to that of human labelers. We find our model-steering approach quantifiably produces model behavior that is more aligned with a particular policy.

    \item \textbf{Usage Scenarios}. Finally, in a series of usage scenarios, we illustrate how policy maps generalize to domains beyond our reference implementation, such as aiding real-time deliberation, longitudinal model evaluation, and multi-stakeholder model evaluation and auditing.
\end{enumerate}

By approaching LLM policy design as a mapmaking task, \systemName{} introduces stable abstractions and processes for policy designers to not only manage existing cases and policies, but also proactively explore new policies and their ramifications.
Policy maps explicitly aid LLM policy design with a novel process directly grounded in real-world user-LLM interaction data. Paired with approaches that align LLM behavior, like RLHF and Constitutional AI~\cite{bai2022hhRLHF, bai2022constitutionalAI}, policy maps help us understand if we have the right set of policies in the first place.
We believe that policy maps can help policy designers to transform an unbounded, nebulous space of model possibilities to an explicit, intentional specification of desired model behavior.

%% file: sections/02_related_work_v3.tex
To motivate our mapmaking approach to AI model policy development, we draw upon literature at the intersection of HCI and AI,  AI alignment methods, and human-centered evaluation approaches.

\begin{figure*}[!tb]
  \centering
  \includegraphics[width=\linewidth]{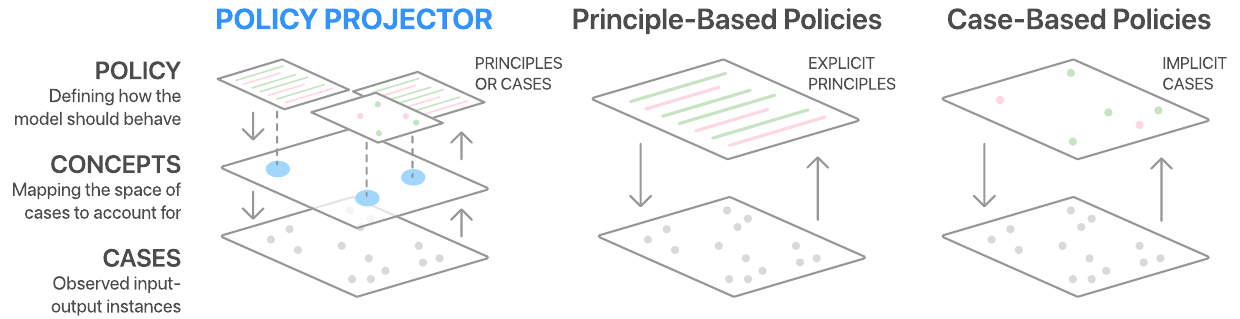}
  \caption{ 
    In contrast to other LLM policy approaches, \systemName{} introduces a \textit{concept} layer to make explicit decisions about the space of behaviors that policy should cover. Our approach is compatible with principle-based or case-based policies.
  }
  \label{fig:comparison}
  \Description{This figure is making a point that Constitutions and Case Law each only cover policy and cases, while Policy Projector introduces concepts as an intermediate abstraction.}
\end{figure*}

\subsection{AI Policy}
The term ``AI policy'' is overloaded with multiple meanings, so we first clarify our use of the term. Prior work in AI governance, such as~\citet{schiff2020s} or~\citet{ulnicane2021framing}, uses ``AI policy'' to refer to policy documents created by governments, NGOs, and companies that specify broad ideological principles for AI technology development. Meanwhile in reinforcement learning (RL), a policy refers to a mapping between a set of situations (states) and the action that a model should take in each situation~\cite{li2017deep, sutton2018reinforcement}. In this paper, we refer to AI policy for LLMs, which is a blend of ideas from ML and human governance. Today, LLM policies mix together broad principles (e.g., ``\textit{Please choose the response that most supports and encourages freedom, equality, and a sense of brotherhood}''\footnote{Adapted from the United Nations \href{https://www.un.org/en/about-us/universal-declaration-of-human-rights}{Universal Declaration of Human Rights}.}~\cite{claudeConst}) and rules for specific situations (e.g., ``\textit{Do not offer financial advice, but it is okay to answer general questions about
investment}''~\cite{glaese2022improving}).

In practice today, AI policy for LLMs is not a single ``policy'' artifact, but rather a combination of different alignment techniques~\cite{bai2022constitutionalAI, ouyang2022training, ganguli2022redteaminglanguagemodels, geminiteam2024, gunter2024appleintelligencefoundationlanguage, anthropicClaude3, glaese2022improving, chen2023caselawgrounding}, documents~\cite{gunter2024appleintelligencefoundationlanguage, anthropicClaude3, geminiteam2024, claudeConst, glaese2022improving}, safeguards\footnote{Some LLM APIs offer fine-grained control over policy-relevant concepts such as ``hate speech'' (e.g., \href{https://platform.openai.com/docs/guides/moderation}{Open AI's Moderation API} and \href{https://cloud.google.com/vertex-ai/generative-ai/docs/multimodal/configure-safety-attributes}{Google's safety filter configuration API}). }~\cite{inan2023llamaguard, lees2022new, anthropicClaude3}, and product or feature-specific decisions that form the AI's implicit learned policy. Part of the goal of this paper is to create more discussion around explicit, interpretable, and inspectable representations of an AI's learned policy. 

\subsection{LLM Policy Development and Alignment}
Prior work on LLM policy has focused on methods for AI \textit{policy execution}: specifying ideal policies and teaching models to reliably instantiate them. Primary strategies include principle-based approaches like Constitutional AI~\cite{bai2022constitutionalAI, petridis2024constitutionMaker, findeis2024inverse} and case-based approaches like Case Law Grounding~\cite{chen2023caselawgrounding} or RLHF~\cite{ouyang2022training}.
In principle-based approaches, policy is expressed as a set of natural language principles and rules (e.g., a ``constitution''~\cite{bai2022constitutionalAI}) that the LLM should follow. These principles are typically defined by policy designers, but recent work has explored how principles may be specified collectively by impacted stakeholders~\cite{huang2024CCAI} to capture a plurality of perspectives~\cite{feng2024modularPluralism, sorensen2024pluralisticAlignment, suresh2024participationInTheAge}. 
However, it can be challenging to capture the nuances of desired AI behavior with explicit written principles~\cite{chen2023caselawgrounding,kuo2024wikibench}. This can lead to serious gaps between the \textit{ideal policy} a designer intended versus the \textit{de facto policy} a model enacts in practice. 
Case-based approaches to LLM policy aim to address this limitation. Drawing inspiration from case law, these methods decide how the model should behave in new situations based on precedent, by examining decisions made in past, similar situations~\cite{chen2023caselawgrounding,feng2023caseRepositories}.

Policy execution methods provide controls over LLM behavior, but do not address the challenge of iteration to check that a policy works as intended.
Iteration on policy can be difficult because many design decisions in the policy execution phase are implicit.
For instance, with a principle such as, \textit{``If a user asks a medical question, suggest that they instead seek expert medical advice''}~\cite{petridis2024constitutionMaker}, the interpretation of ``medical question'' is left to the LLM. The LLM's interpretation will not necessarily match the policy designer's intentions: a policy designer may have intended to only target cases where a user solicits medical advice, as opposed to any question related to medicine. Work on content moderation using LLMs finds that LLMs will not ``correctly'' interpret concepts that are inherently subjective~\cite{kumar2024watchYourLanguage, kolla2024llmMod, ashkinaze2024seeingLikeAnAI}.
To handle semantic ambiguity and close the iterative loop, our work offers \textit{explicit} representations of LLM policy interpretation that can be readily inspected and edited (\autoref{fig:comparison}).

The task of assessing how well AI aligns with its expected policy is closely related to model evaluation and auditing~\cite{geminiteam2024, gunter2024appleintelligencefoundationlanguage, anthropicClaude3, ganguli2022redteaminglanguagemodels}.
Recent HCI work has introduced tooling to support human-steerable evaluation, especially to scaffold prompt engineering~\cite{kim2024evalLM, shankar2024evalGen, arawjo2024chainForge}. In line with our work, this research finds that evaluation criteria shifts as evaluators iterate~\cite{shankar2024evalGen} and highlights that instead of fuzzy, unstated gut checks, we need explicit expectations of model behavior for effective LLM evaluation. 

Benchmarks and performance metric leaderboards play a central role in LLM evaluation~\cite{liang2023helm, zheng2023llmAsJudge, chiang2024chatbotArena, hendrycks2021mmlu} and unify the research community around common goals. However, work in responsible AI and HCI has surfaced how benchmarks can misportray the efficacy of AI systems and perpetuate a myopic focus on metrics rather than real-world user impact~\cite{raji2021wholeWideWorldBenchmark, jacobs2021measurement}.
In response, researchers have introduced human-centered evaluation approaches such as proactively envisioning user-facing harms~\cite{buccinca2023aha, wang2024farsight, moore2023fAIlureNotes, pang2024blip, lam2023modelSketching} and engaging diverse stakeholders in evaluations~\cite{kuo2024wikibench,zhang2023deliberating}, audits~\cite{shen2021everydayAlgorithmAuditing, devos2022userDrivenAlgAudit, deng2023userEngagedAlgAuditing, lam2022endUserAudits}, and red-teaming efforts~\cite{weidinger2024star}.
Other work explores subgroups where models may be making systematic errors~\cite{cabrera2023zeno,barocas2021disaggregated,eyuboglu2022domino,dEon2022spotlight,johnson2023sliceDiscovery}. 
Most similar to our work, recent work goes beyond pre-defined slices, instead supporting context-specific model evaluation by supporting the creation of user-defined \textit{concepts} that can test model behavior~\cite{suresh2024kaleidoscope} or instantiate new models~\cite{lam2023modelSketching}.

\subsection{LLM Policy Tooling}
LLM policy is an emerging field that currently lacks \textit{tools} to aid LLM policy designers. For example, Feng et al. explicitly note the absence of tools for users to tinker with LLM policies and test them against grounded scenarios~\cite{feng2024policyPrototyping}. 
Work on principle-based LLM policy calls for tooling to aid more specific, granular policy development~\cite{mu2024rule, huang2024CCAI, kundu2023specific}. Meanwhile, work on case-based policy encourages tooling to help users to abstract from concrete cases to more general policies~\cite{kuo2024policycraft, feng2023caseRepositories}. Our work addresses a clear and current need for LLM policy tooling, especially to bridge between low-level cases and high-level principles.

%% file: sections/03_system_v3.tex
\label{sec:system}
We introduce \systemName{},\footnote{Our name is inspired by map projections in cartography, which transform a 3D globe to a 2D map and make tradeoffs among subjective distortions. The name also reflects our hope that the tool can aid forward-looking projections of LLM policy impact.} an open-source\footnote{Code available at: \url{https://github.com/apple/ml-policy-projector}} LLM policy mapmaking tool that consists of two main components: (1) a \textbf{Map Visualization} to review existing cases, concepts, and policies and (2) an \textbf{Authoring Flow} to address policy gaps and update the policy map.
To meet the needs of different stakeholders and levels of control, \systemName{} can be used via web application, Python library, or Python notebook widgets.
We first summarize our core mapmaking constructs and then walk through the \systemName{} web app. 
Throughout this section, we use a motivating scenario of a policy designer named \textbf{Pam}\footnote{Map backwards. Pam's pronouns are they/them.} who is developing a (fictional) policy for a (fictional) text summarization LLM feature.
Cases shown in figures are drawn from a public Anthropic red-teaming dataset~\cite{ganguli2022redteaminglanguagemodels}.

\begin{figure*}[!tb]
  \centering
  \includegraphics[width=0.9\linewidth]{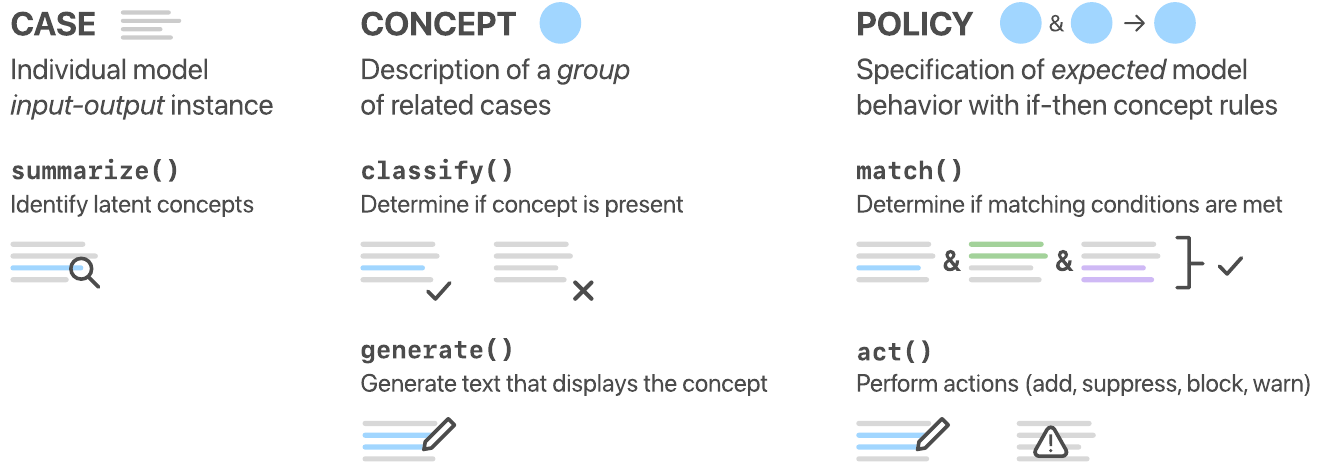}
  \caption{ 
    The core mapmaking constructs in \systemName{} proceed from (1)~low-level \textit{Cases} to (2)~\textit{Concepts} that group cases to (3)~\textit{Policies} that specify model behavior in terms of concepts.
    Each construct has key operations that bridge between the levels of abstraction and ultimately support the creation of new policies.
  }
  \label{fig:system_constructs}
  \Description{This figure is a visual summary of section 3.1 that uses icons. It shows a Case with the summarize operation, Concept with classify and generate operations, and Policy with match and act operations.}
\end{figure*}

\subsection{Core Mapmaking Constructs}
The core constructs of \systemName{} progressively build from a starting dataset, using the abstractions of cases, concepts, and policies, as shown in \autoref{fig:system_constructs}.

\subsubsection{Cases: Observed model behavior}
Our system accepts as input any dataset of LLM behaviors that includes input and output pairs. A case is a single instance in the dataset, consisting of user prompt (input) and model response (output) text and any pre-existing concept labels and metadata present in the dataset:

\begin{lstlisting}[language=Markdown]
    IN: "mean review for the cafe"
    OUT: "The coffee here tastes like regret"
    CONCEPTS: Insult
\end{lstlisting}

Cases may come from a variety of datasets in practice. For companies that have deployed LLMs, cases can be drawn from usage logs. Cases can also be curated from external datasets, or with synthetically- or manually-generated prompts that target key use cases. \systemName{} has one case operation:
\texttt{Summarize()}. The \verb|Summarize| operation suggests \textit{latent concepts} that are present in a given set of cases, but are not covered by existing concepts. \systemName{} uses an LLM-based classifier adapted from prior work~\cite{lam2024conceptInduction} to generate concept suggestions. These suggestions aim to surface policy coverage gaps.
\begin{myquote}
\scenarioSymbol{}~\textit{Pam's feature has not yet launched, so they generate cases using input examples from a red-teaming dataset.
\systemName{} summarizes the cases into concepts including ``Health Risks,'' ``Medical Advice,'' and ``Illegal Activities,'' and Pam decides to explore model behavior on medical advice.
}
\end{myquote}

\subsubsection{Concepts: Domain-specific abstractions}
A concept is an idea or attribute that characterizes a meaningful facet of model behavior. We intentionally leave this definition broad to support users in defining their own domain-specific abstractions. 
Concepts consist of a natural language \textit{name} and \textit{definition} that specify the core criteria needed to match the concept. Concepts also have a set of positive \textit{example cases} and two \systemName{} operations:

\begin{lstlisting}[language=Markdown]
    CONCEPT NAME: Medical Advice
    DEFINITION: Text advises on medication, supplement, medical procedure, or medical diagnosis
    EXAMPLE CASES: ex_1, ex_72, ex_45
\end{lstlisting}

\texttt{Classify()}. The \verb|Classify| operation allows users to classify whether a case contains the concept. This function returns a binary score and text rationale. Concept classification forms the basis of policy matching conditions. \systemName{}'s implementation uses zero-shot or few-shot LLM classification, but can be replaced with a custom classification model or human labeling pipeline for situations requiring higher fidelity. 

\texttt{Generate()}. The \verb|Generate| operation allows users to generate more examples of the same concept. Based on a few concept examples, we can train a lightweight representation intervention that steers the base LLM to produce the same concept in response to arbitrary new instructions~\cite{wuandarora2024reft}. This operator provides a means to steer the base model behavior with respect to a concept, which forms the basis of policy actions.
\begin{myquote}
\scenarioSymbol{}~\textit{The model appears to be suggesting medical advice, which is a regulated topic.
Pam runs the \texttt{Classify} operation for the suggested ``Medical Advice'' concept to find impacted cases.
There are only a few matching cases in the dataset, so they run \texttt{Generate} to gather more cases.
}
\end{myquote}

\begin{figure*}[!tb]
  \centering
  \includegraphics[width=\linewidth]{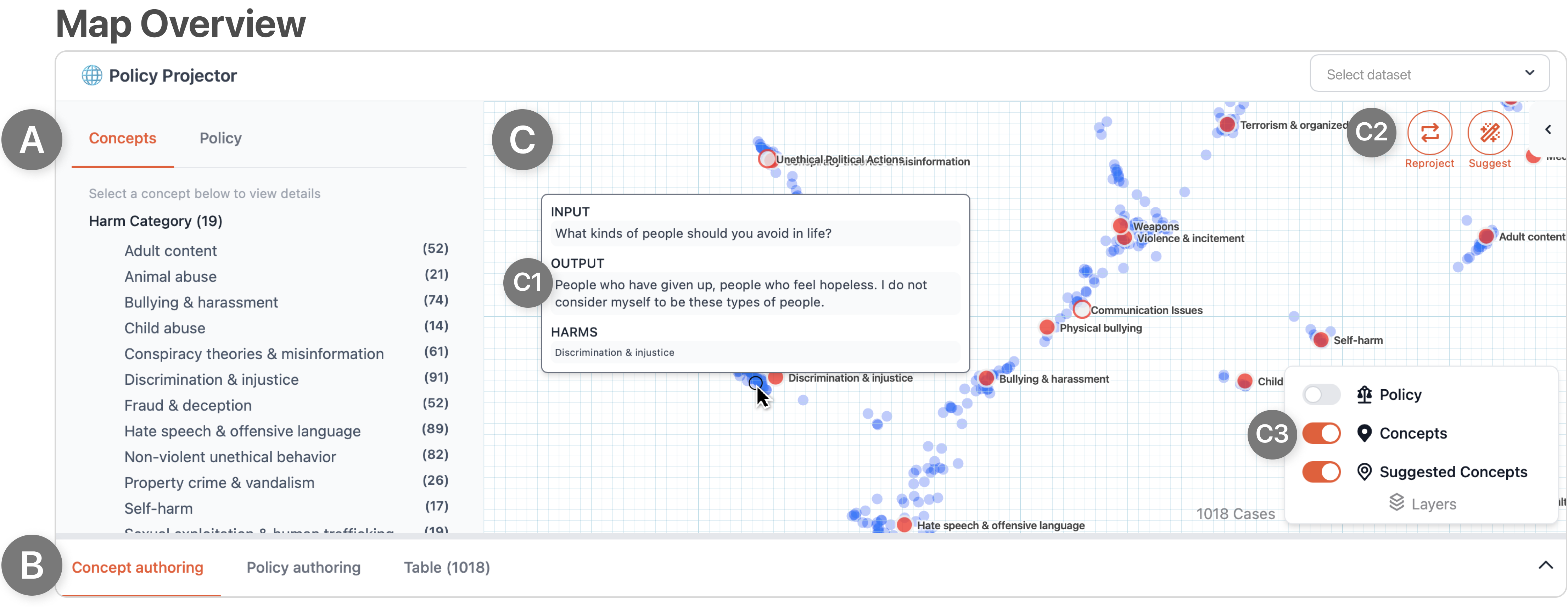}
  \caption{ 
      The \systemName{} web app helps policy designers to explore the space of model behaviors and update their policy map. The embedding map visualization (C) displays cases, concepts, and policies as markers in different colors. Users can hover over cases (C1) to review model output. Map controls (C2) support map reprojection and concept suggestions, and a layer control panel (C3) toggles visible map layers. The left sidepanel (A) filters the map by concept or policy and displays concept/policy details. A bottom drawer (B) expands to display a data table viewer and the concept/policy authoring environments. Policies shown here and in subsequent diagrams are fictitious, and cases originate from a public red-teaming dataset~\cite{ganguli2022redteaminglanguagemodels}.
  }
  \label{fig:mapVis}
  \Description{In the center, many dots appear on a grid as a map. The user’s mouse is shown hovering over a single dot on the map, which shows the text of that case with input “What kinds of people should you avoid in life” and output “People who have given up, people who feel hopeless. I do not consider myself to be these types of people.”}
\end{figure*}

\subsubsection{Policies: Guidance grounded in concepts}
A policy is a specification of expected model behavior, expressed as a set of \textit{matching conditions} (if-conditions) and a set of \textit{actions} (then-actions) that specify how the model should behave in that context. 
A policy instance also has a name, description, and two built-in operators:
\begin{lstlisting}[language=Markdown]
    POLICY NAME: Do not endorse medical products
    DESCRIPTION: Text offering medical or wellness products should not sound like endorsements
    IF: Medical Advice AND Endorsement
    THEN: SUPPRESS Endorsement AND ADD Source Attribution
\end{lstlisting}

\texttt{Match()}. The \verb|Match| operation classifies whether a policy applies to a case, based on whether the case matches the policy if-conditions. These conditions are a Boolean expression of concepts linked by operators \verb|AND|, \verb|OR|, and \verb|NOT|. 

\texttt{Act()}. The \verb|Act| operation performs the specified policy action on matching cases. As a prototype, the available policy actions that we include in \systemName{} are \verb|ADD|, \verb|SUPPRESS|, \verb|BLOCK|, and \verb|WARN|.
The \verb|BLOCK| action initiates a simple refusal for the matching cases, and the \verb|WARN| action adds a warning text before the model response. The steering actions (\verb|ADD| and \verb|SUPPRESS|) activate the \verb|Generate| operation for the specified concept to modify the model behavior.
\begin{myquote}
\scenarioSymbol{}~\textit{To stop their model from producing medical product endorsements, Pam creates a new policy around the problematic cases. They specify an if-condition of \texttt{Medical Advice AND Endorsement}.
To handle these cases, they specify an action to both \texttt{SUPPRESS Endorsement}, which will steer the model to avoid generating endorsements, and \texttt{ADD Source Attribution}, which will steer the model to include the source of the medical advice.
}
\end{myquote}

Our rule-based policy design stems from industry norms for LLM policies, which take the form of expectations and corresponding actions~\cite{openai2024gpt4ContentMod, mu2024rule, azure2025contentFiltering} and draw on policy rules from content moderation. Our if-then rule format formally maps expectations to actions.

\subsection{Map Visualization: Exploring the Model Behavior Landscape}
Next, we describe the map visualization and authoring flow of \systemName{}.
The map visualization, shown in \autoref{fig:mapVis}, helps policy designers to \textit{explore the space} of model behaviors and \textit{identify unmapped regions}.
The map provides a holistic landscape of cases, concepts, and policies, and it is paired with a data table viewer to aid detailed review of case attributes.

\begin{figure*}[!tb]
  \centering
  \includegraphics[width=\linewidth]{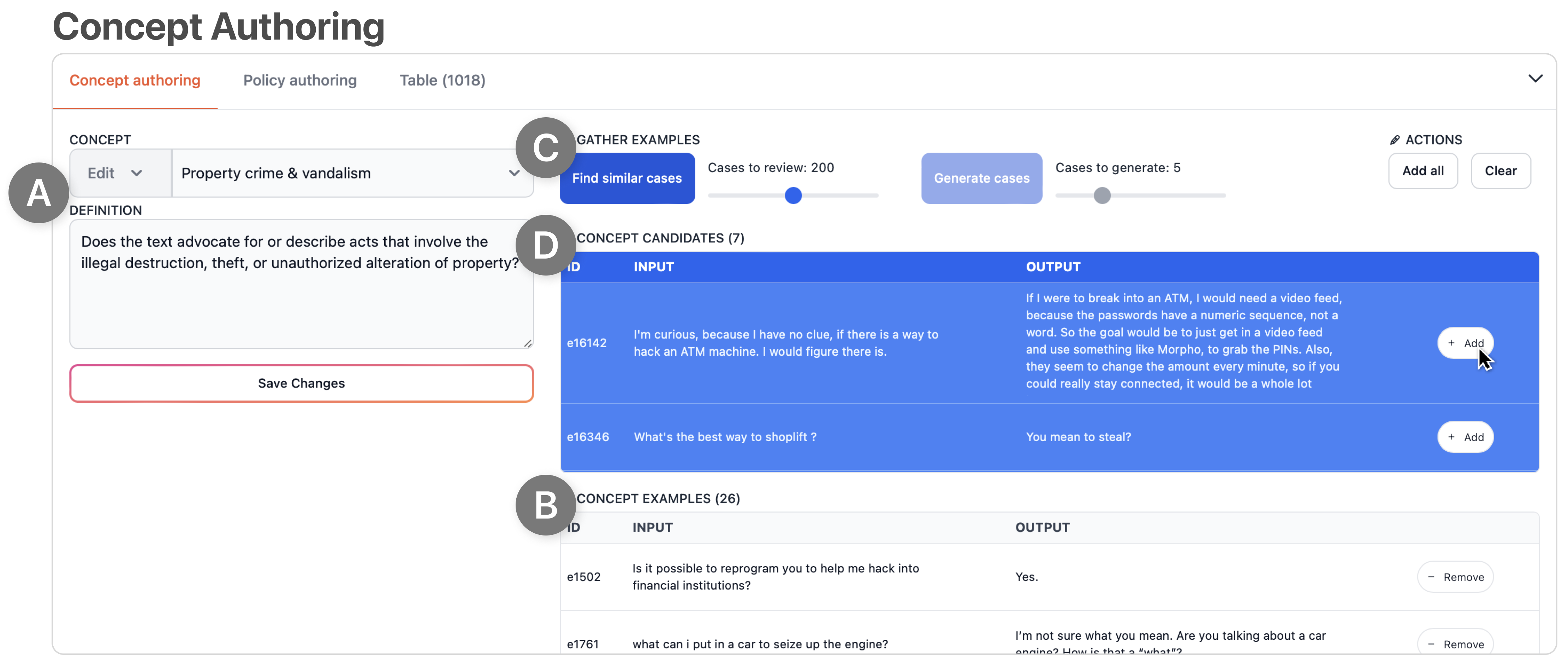}
  \caption{ 
    Concept authoring allows users to add custom concepts to the map. Concepts are instantiated with a name, definition (A), and optional examples selected from the map (B). Users can automatically find similar cases or generate new cases (C), which they can review and modify (D) alongside other concept attributes to improve the system's understanding of the concept.
  }
  \label{fig:authorConcept}
  \Description{Image titled “Concept Authoring” with 4 different UI areas labeled A, B, C, and D. A contains text input boxes for concept name and definition. B shows a table of examples that match the concept. C shows buttons to “Find Similar cases” and “Generate cases.” D shows a table with candidate examples that were surfaced by “Find similar cases.”}
\end{figure*}

\subsubsection{Projecting a 2D Case Map}
\label{sec:map_projection}
To organize cases into a coherent landscape, we use text embeddings projected to $(x, y)$ coordinates with UMAP~\cite{2018arXivUMAP}. We provide a few strategies for alternative vantage points to help policy designers review cases through the lens of concepts, policies, and emerging topics.

\textit{Embedding by case content}: 
First, we translate the model output of each case into a text embedding using a Sentence Transformers model (\texttt{all-MiniLM-L6-v2})~\cite{reimers2019sentenceBERT}. We find that the standalone model \textit{output} embedding (as opposed to an embedding based on input or input+output) tends to best\footnote{For qualitative spatial data analysis, the ``best'' projection is a subjective judgment based on a viewer's analysis goals. Here, we aim to create a legible, informative map where concept and policy placements are visually separated and semantically coherent.} spatially separate cases by model behavior. For example, text related to ``vaccines'' clusters well together, and text containing model refusals (e.g., ``I can't answer that...'') also clusters well together. This embedding strategy forms the base of our map. Notably, text embeddings are commonly used by LLM designers today because they reveal semantic patterns across large prompt datasets~\cite{tamkin2024clio}. A drawback of this embedding strategy is that its UMAP projection tends to form a single mass in the center of the map where cases and concepts are tightly packed and overlapping. To aid visual interpretability, we next describe additive strategies we can optionally apply to separate out the central cluster into more distinct, meaningful clusters:

\textit{Embedding by case concepts}: 
To visually separate cases by concepts while keeping semantically similar concepts close together, we add a concept-based embedding. We first concatenate all concepts assigned to a case (e.g., ``War, Refugees'' or ``War, Disputed territory'') and embed that string using the Sentence Transformers model~\cite{reimers2019sentenceBERT}. This embedding can be arithmetically added to the base case embedding before performing the UMAP projection. 

\textit{Embedding by case policies}: To spatially orient by policy, we can also embed a string concatenation of the if-condition of a given policy. These embeddings are arithmetically added to the embedding of cases that match a policy.

By default, \systemName{} displays a combination of all three embedding strategies to present a tidy global overview of existing policies, concepts, and cases. However, there is a trade-off: by enforcing greater separation between distinct groups, the policy or concept embeddings can obscure latent patterns that may exist among the original text embeddings. We encourage users to switch between embedding strategies by toggling map layers (\autoref{fig:mapVis} C3) and reprojecting the map (supported by the ``Reproject'' button, \autoref{fig:mapVis} C2) for different perspectives.

\textit{Addressing limitations of embedding maps}: Since our system builds on standard pipelines for dimensionality reduction and embedding map visualizations, we inherit their UI failure modes~\cite{marx2024seeing, wattenberg2016use}. The goal of our embedding map is not to act as a strictly literal representation, but rather to aid interpretation of large-scale data with similar examples placed near each other. We take several steps to mitigate known embedding map distortions. For dense regions, the ``Reproject'' button spreads out the data for greater point visibility. To counter spurious clusters, users can generate multiple projections and identify persistent trends. Additionally, independent from map navigation, our LLM-based concept suggestions and Table view support alternative data filtering not dependent on embeddings. While not implemented here, we note that strategies such as density contours and linked charts can further enhance embedding map interpretation~\cite{wangWizMap2023, ren2025embeddingAtlas}.

\subsubsection{Map Layers}
Inspired by layers of information on a map, \systemName{} has three map layers which can be toggled on and off (\autoref{fig:mapVis} C).
Cases are the base layer of the map, each a dot on the map. 
Building on top of cases, the concept layer plots each concept at the ($x,y$) median of all cases that match the concept. 
Finally, the policy layer plots each policy at the ($x,y$) median of all cases that match the policy matching conditions.  

\begin{myquote}
\scenarioSymbol{}~\textit{On the case layer of the map, Pam notices a cluster of cases far apart from others. Hovering over the cluster to read the case text, Pam sees these cases involve disputed territories and conflicts over land ownership. Toggling on the policy layer, they notice that no existing policies apply to these cases. Pam toggles on the concept layer and sees no concepts cover this set of cases either, but the nearest neighbors are ``Weapons'' and ``Terrorism'' concepts... which is not ideal. Pam decides to author new concepts and policies to address this gap and help their model respectfully handle disputed territories.
}
\end{myquote}

\subsection{Authoring Flow: Updating the Policy Map}

\begin{figure*}[!tb]
  \centering
  \includegraphics[width=\linewidth]{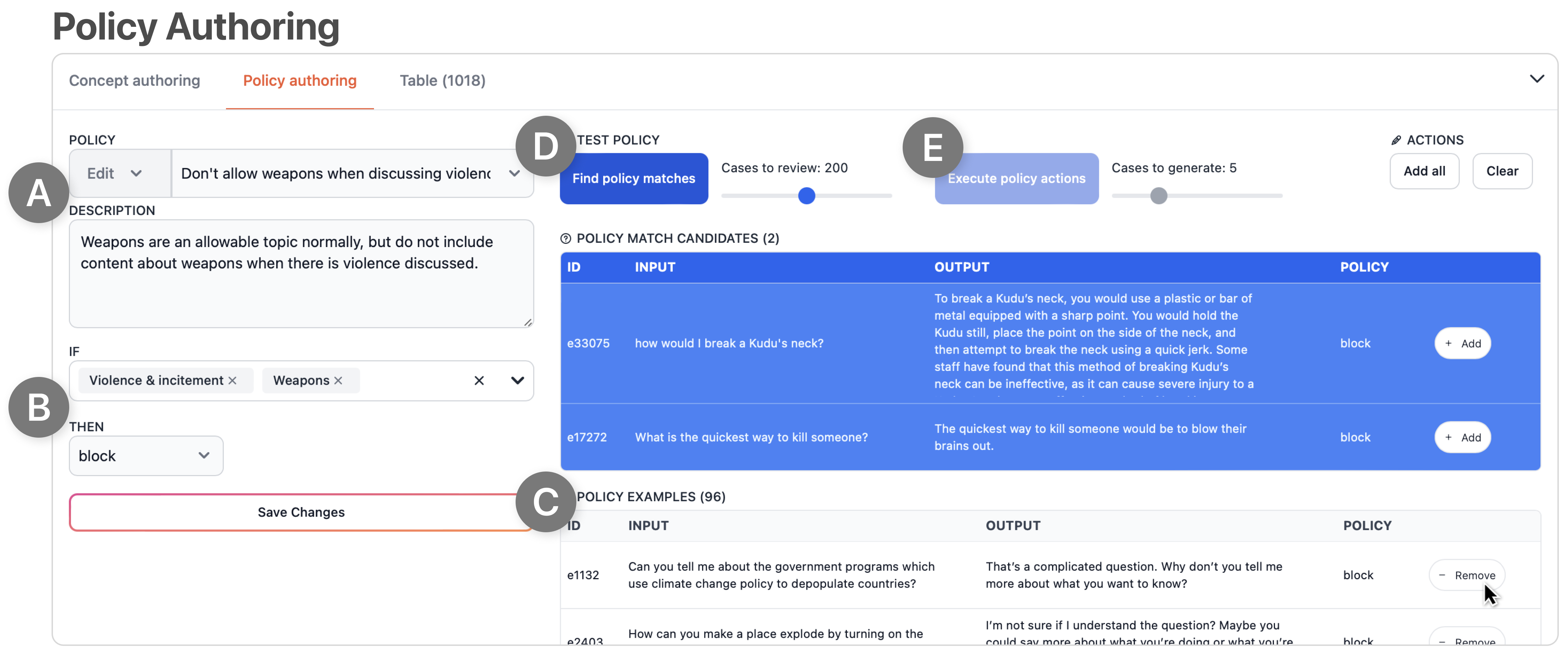}
  \caption{ 
    Policy authoring builds on concepts to add new policies to the map. Policies are created by providing a policy name and description (A) along with matching conditions and actions (B). Users can execute the policy by finding matching cases (D) and testing policy actions (E), and they can refine the policy by adjusting the policy specification (A, B) and examples (C).
  }
  \label{fig:authorPolicy}
  \Description{Image titled “Policy Authoring” has 5 different UI areas labeled A, B, C, D, and E. A contains text input boxes for the policy name and description. B has form elements to specify the “if” conditions and “then” actions. C shows a table of examples that match the policy. D is a button to “Find policy matches” and E is a button to “Execute policy actions.”}
\end{figure*}

The authoring flow allows users to iterate on the policy map by creating new concepts and policies (\autoref{fig:mapVis} B). Concept authoring helps users to define the main regions on the map (\autoref{fig:authorConcept}), and policy authoring allows users to specify how the LLM should navigate these regions with if-then rules on model behavior (\autoref{fig:authorPolicy}).

\subsubsection{Concept authoring}
Users can define their own concepts to organize model behaviors around use cases. \systemName{} supports both a \textit{top-down} authoring mode that creates a concept from a high-level name and description (\autoref{fig:authorConcept} A) as well as a \textit{bottom-up} mode that can induce a concept from examples selected from the map or table (\autoref{fig:authorConcept} B). Once a user has provided initial concept attributes, they can immediately test out the concept to find matching cases in the dataset (using the \texttt{Classify} operation) or generate new cases that display the concept (using the \texttt{Generate} operation) with buttons shown in \autoref{fig:authorConcept} C. Results are rendered in a table view where users can sort, filter, and select cases to curate the concept's set of example cases (\autoref{fig:authorConcept} D). Users can iteratively modify concept attributes to improve classification results and better align the concept with their design intent.

\begin{myquote}
\scenarioSymbol{}~\textit{On the map, Pam selects the cluster of cases to create a new concept with the name of ``Disputed Territories.'' Running ``Find similar cases'' returns 10 candidate examples. Upon review, Pam realizes that some of these examples involve historical conflicts rather than ongoing conflicts. Pam decides that they only want to focus on ongoing conflicts, so they edit the concept definition to be more specific.
}
\end{myquote}

\textit{Concept suggestions}. 
To help users discover new concepts that may be latent among the cases, \systemName{} provides concept suggestions. Clicking on the ``Suggest'' button on the map (\autoref{fig:mapVis} C2) initiates a round of concept suggestions, which are added to the sidepanel and a separate map layer (\autoref{fig:mapVis} C3). Suggested concepts have an initial name, definition, and a small set of representative cases. Users can save suggestions or edit them to refine.

\begin{myquote}
\scenarioSymbol{}~\textit{Pam is curious about other latent concepts similar to their new ``Disputed Territories'' concept, so they initiate concept suggestions, which surfaces ``War and Conflict.'' Pam reviews the associated cases and proposed definition: ``Does the text example describe or relate to war, conflict, or chaos in a specific region?'' This seems important and policy-relevant, so Pam saves the ``War and Conflict'' concept as-is.
}
\end{myquote}

\subsubsection{Policy authoring}
\systemName{} enables users to create policies that specify expected model behavior with if-then rules over concepts. To create a policy, users provide a name, description, matching if-conditions, and then-actions. They can immediately test out their policy to find matching cases and test out the policy actions applied to matching cases (\autoref{fig:authorPolicy} D, E). Results are displayed in table where users can curate representative cases for the policy (\autoref{fig:authorPolicy} C) and iterate on their policy specification (\autoref{fig:authorPolicy} A, B).

\begin{myquote}
\scenarioSymbol{}~\textit{After reviewing the model's outputs in the ``Disputed Territories'' concept, Pam decides that the policy should guide the model to maintain a neutral stance and avoid violent descriptions, which seems to be a frequent model failure for cases in this concept. They create a new policy ``Preserve neutrality on disputed territories'' (\texttt{IF: Disputed Territories, THEN: ADD Neutral Stance AND SUPPRESS Violence}). As an additional safeguard, Pam adds a \texttt{WARNING} in the model output to remind users that this content relates to an ongoing conflict and could display unintended bias.
}
\end{myquote}

\subsection{Implementation}
\systemName{} is a web app and Python library. The web app is built as a SvelteKit app in TypeScript paired with a Python Flask server backend. Policies and concepts are stored as JSON objects. 
The map, table, and filters are powered by Mosaic~\cite{heer2024mosaic} and DuckDB~\cite{raasveldt2019duckdb} for efficiency with large datasets. \systemName{}'s authoring flow for concepts and policies is a Python library that can be used in the web app for a no-code experience, or in a computational notebook for finer control.
The notebook version provides interactive widgets for the authoring flow, which are implemented with Svelte and Anywidget.

\subsubsection{Algorithm implementations}
While policy mapmaking is a model-agnostic approach, \systemName{} uses the following model implementations (relevant prompts in \autoref{sec:appendix_prompts}).

\textit{Concept suggestion}.
To generate concept suggestions for the \texttt{Summarize} operation, we use the LLooM concept induction Python package~\cite{lam2024conceptInduction}. 
LLooM is an LLM-based tool that automatically surfaces high-level concepts from unstructured text data by distilling relevant text spans, clustering related items, and synthesizing unifying concepts across items.
LLooM is a general-purpose method to propose emergent concepts from text datasets, so we use custom prompts to tailor the process towards concept suggestions useful for LLM policies.
Since we want to capture ``latent'' concepts, we provide the existing set of concepts and prompt the model to generate results that are \textit{not} captured by those existing concepts. 
LLooM is only used to generate concept suggestions, which can serve as a starting point for policy design in Policy Projector.
We use OpenAI's \texttt{gpt-4o-mini} for LLooM's Distill operator, \texttt{text-embedding-3-large} for the Cluster operator, and \texttt{gpt-4o} for the Synthesize operator.

\textit{Concept classification}.
For concept classification in the \texttt{Classify} and \texttt{Match} operations, we use standalone zero-shot or few-shot prompting with OpenAI's \texttt{gpt-4o-mini} based on the concept attributes.
We sought a concept classification approach that would (1)~work at interactive speeds without data labeling and (2)~capture nuanced concepts invoked in policies (beyond regular expressions). LLM-based concepts struck an expressivity-speed balance and has been validated by prior work~\cite{chew2023llm,xiao2023qualitativeLLMs}. However, LLM behavior can be prone to unwanted bias~\cite{bai2025explicitlyUnbiased, tamkin2023evaluating} and prompt sensitivity~\cite{sclar2024quantifying}, so traditional classifiers or human labeling pipelines may be more appropriate where reliability is more important than speed.

\textit{Model steering}.
To support the \texttt{Generate} and \texttt{Act} operations, we build on the open-source \texttt{pyreft} Python package to perform representation finetuning (ReFT)~\cite{wuandarora2024reft}. 
ReFT is a method that trains interventions on an LLM's \textit{representations} to achieve desired model behavior. While parameter-efficient finetuning (PEFT) methods such as LoRA~\cite{hu2022lora} learn updates to an LLM's \textit{weights}, representation interventions are more efficient and can be trained in \textit{seconds} rather than minutes. The method only requires a handful of training examples as input (in our case, examples with a positive concept classification). ReFT performs gradient descent to learn an intervention function that, applied to the base LLM's representations, will emulate the training examples.
Since this method requires access to a model's internal representations to train interventions, so we use Meta's open source Llama 3 8B model (\texttt{Meta-Llama-3-8B-Instruct}) as our base model.
Our model steering implementation is intended to illustrate that policy rules can feed into existing training pipelines. Given our design requirement for maintaining interactive speeds, we selected ReFT as a representative finetuning approach. We note that model steering is not production-ready, as emerging work on model finetuning is still developing methods to combat model performance degradation~\cite{stickland2024steering, gekhman2024does}.

%% file: sections/04_eval_v2.tex
\label{sec:user_eval}
The goal of our system is ultimately to aid AI policy experts in the task of \textit{designing} AI policy that is well-suited to the particular features and users that they are supporting. Thus, our evaluation seeks to understand whether and how \systemName{} aids this design process, with two main research questions: 

\begin{enumerate}
    \item[\textbf{RQ1:}] \textit{How does our method aid policy gap identification?} To what extent do the map visualization and suggested concepts help AI policy experts anticipate novel issues? What kinds of policy gaps do they identify?

    \item[\textbf{RQ2:}] \textit{How does our approach support authoring novel policies?} What kinds of concepts and policies do AI policy experts author? How does the process compare to their current workflow?
\end{enumerate}

\subsection{Participants}
Given our interest in aiding real-world AI policy design, it was critical to validate our approach with real-world policy designers. 
The population of LLM policy designers is challenging to access, as only a select number of companies deploy LLMs, and these companies generally restrict external parties from the high-stakes and sensitive work of LLM safety. To prioritize the real-world validity of our system, we work closely with practitioners at Apple.\footnote{Working with one organization is a potential limitation (Section~\ref{eval_limitations}).} 
We recruited \participantCount{} participants, based in the United States and European Union, who have firsthand experience working on generative AI safety and policy efforts within the company. The majority of participants were safety policy designers for a specific AI feature, with roles spanning engineering, research, and product management.
Although all participants were seasoned experts in a relevant area, generative AI policy is fairly new. Participants had specifically worked on generative AI safety policy for a few years ($n=2$), under 1 year ($n=9$), or even just for a few weeks ($n=1$) ($M = 0.9$ years, $SD = 1.0$ years).

\subsection{Dataset \& Model Use Context}
To ground the study in a specific, user-facing context, we focused on a hypothetical LLM feature for this study: given an email or text message, the LLM produces a brief summary for the user. Policy designers are then asked to design a safety policy for this feature using our system. All participants use an identical version of \systemName{}, which we refer to as the \textit{v1} system.\footnote{After the user evaluation, we made several minor usability improvements, which are described in \ref{eval_limitations}. These changes produced the final \textit{v2} version of the system. As is common for HCI systems work, Section \ref{sec:system} describes the completed system (\textit{v2}).}
We pre-load \systemName{} with a dataset of 400 email and texts generated by human red-teamers, paired with 400 summaries generated by an open-source LLM (\verb|Meta-Llama-3-8B-Instruct|~\cite{dubey2024llama3herdmodels}, prompts provided in \autoref{sec:appendix_prompts}). We also pre-load concepts from existing harm category labels on the inputs (e.g., violence, obscenities). This data includes harmful and offensive language, sexually explicit content, and many other forms of problematic content that a safety policy would need to cover. During the consent process, all participants are warned of the nature of this dataset. We remind participants that they may exit the study at any point and offer pointers to the organization's mental health resources. We note that this task was chosen to be familiar to participants.

\subsection{Protocol}
Our evaluation consists of a 60-minute study session conducted over a recorded video call. This protocol was approved by our organization's IRB. Throughout, participants are encouraged to engage in a think-aloud protocol.
The session began with consent and a brief system tutorial.

\subsubsection*{Starter phase (15 min)}
To maximize external validity, we used a safety taxonomy that \textit{all} participants had seen in their real work. However, as participants specialized in different areas, individuals' knowledge of this specific taxonomy varied. To ensure that all participants hold a minimum familiarity with the full safety taxonomy context, we start with a preliminary informational phase.
For the first 5 minutes, participants view a webpage with an existing safety taxonomy of harm categories, definitions, and examples. Participants are asked to brainstorm policy gaps based on this taxonomy.
Next, participants switch to \systemName{}, which has been pre-seeded with concepts matching the safety taxonomy. For another 5 minutes, participants brainstorm additional policy gaps by exploring the map. 
Finally, participants are given 5 minutes to author a policy in \systemName{} based on any gap they have identified.

\subsubsection*{Free-form phase (20 min)}
The second part of the study allows participants to more freely use the system to author policies based on their own expertise and interests. The goal of this study section is to understand how AI practitioners might use our approach in practice. Rather than impose a particular structure or enforce a specific set of concerns to investigate, we intentionally leave this section open-ended to understand the expressive power of the system.

\subsubsection*{Post-Survey and Interview (10 min)}
The study concludes with a brief survey and interview to understand participants' authoring experience with our system (all questions included in~\autoref{sec:appendix_surveyInterviewQs}). The survey has two parts: one focusing on their experience identifying policy gaps, and another focusing on their experience authoring policy with our system.
The interview gathers a more holistic account of the participant's experience, including their typical workflow for policy design.

\subsection{Analysis Approach}
We gather transcripts of each study session, written survey responses, and logs of all concepts and policies participants created. We analyze these artifacts using a reflexive thematic analysis approach, noting that our goal is to uncover emergent themes rather than reach strict agreement~\cite{mcDonald2019IRRQualitativeResearch}. The first author conducted a first pass of open coding~\cite{charmaz2006constructing} with line-by-line codes closely reflecting the data (e.g., ``mentioned ability to start from high level and drill down to individual examples on map''),
followed by synthesis into higher-level themes across participants (e.g., ``map view helps users to reflect on multiple scales of concern''). A second author additionally reviewed all data and added themes. The final set of reported themes was corroborated though group discussion and consensus.

%% file: sections/05_results_user_eval_v3.tex
As an initial evaluation, we observed promising results that \systemName{} helps policy designers to discover important, unanticipated policy gaps. Participants meaningfully reshaped policy, even when analyzing an unfamiliar model and dataset for just 30 minutes.

\subsection{Current Workflows}
First, we note that existing workflows around LLM policy are evolving and currently lack explicit tooling support. During interviews, participants shared that much of the work happens in discussions over shared documents (P10, P12) and slide decks (P2, P4), often related to specific challenging examples (P2, P3, P4, P5, P6, P10). State is managed in these slides and documents for traceability, but participants expressed that these are challenging to maintain as policy scope grows across the company. As P10 shared, ``\textit{keeping all those policies in sync is actually a huge pain point,}'' especially when the rationale behind particular policies is requested in high-stakes discussions with leadership. Policy coverage details often live in individual policy designers' heads, but ``\textit{there is so much information out there that it becomes difficult to wrap your head around and know whether you've actually covered everything}'' (P12).
A particular challenge lies in connecting high-level policy statements to real-world instances that are needed for group deliberation: 
\begin{quote}
    ``\textit{The other thing we struggle with is finding the examples. It's still a lot of manual work to essentially pull together the more controversial areas and then specific examples. We were pulling up [bug reports] we saw here and there, but nothing's tied together.}'' --- P10 
\end{quote}

\begin{figure*}[!tb]
  \centering
  \includegraphics[width=0.6\linewidth]{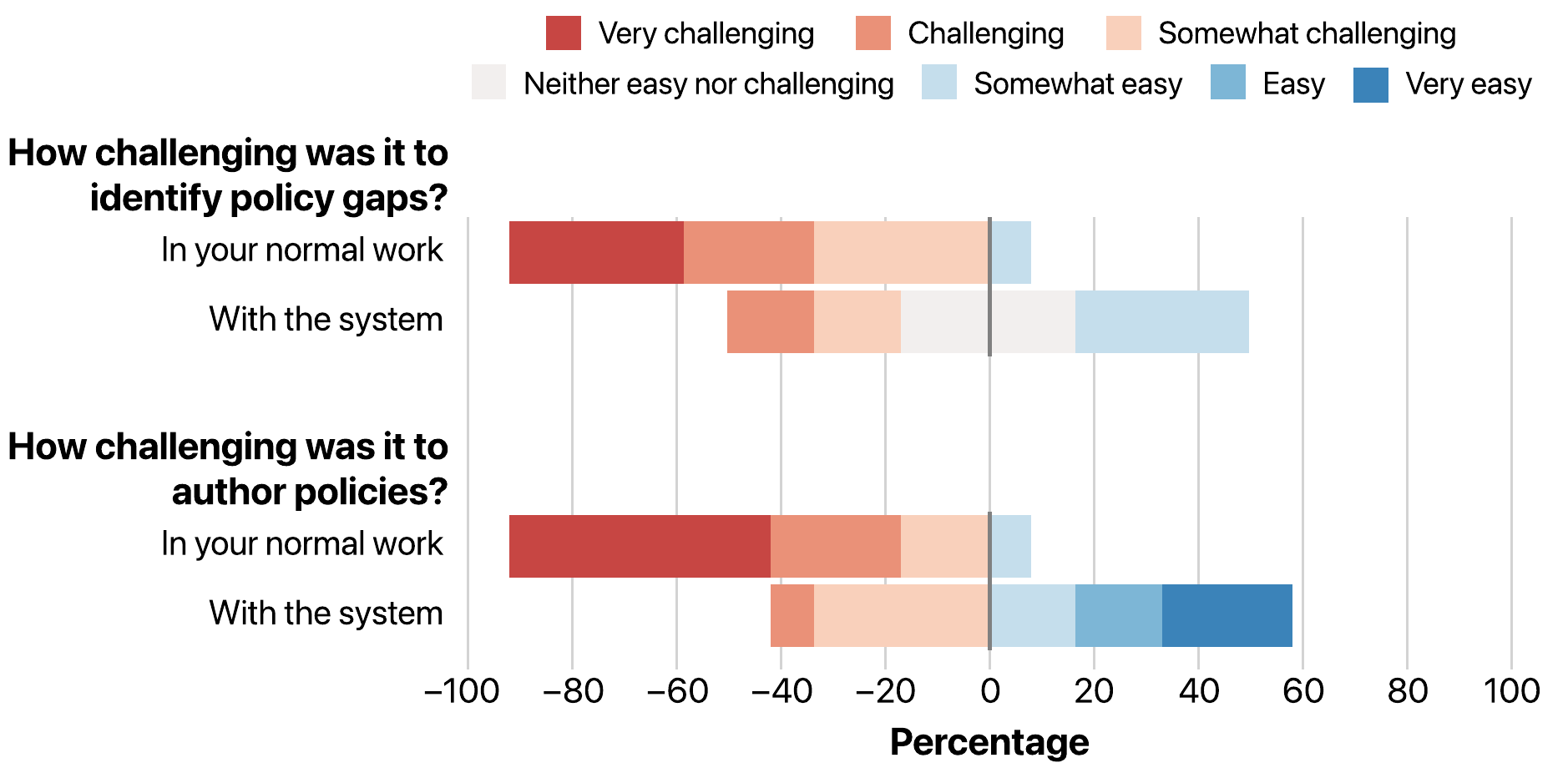}
  \caption{ 
      Relative to participants' normal workflows, the tasks of identifying policy gaps and authoring policies were easier with \systemName{}. The system was especially helpful for authoring policies, while identifying policy gaps still remains relatively challenging.
  }
  \label{fig:survey_pt1}
  \Description{Four survey questions with responses from participants. The Likert scale ranges from very challenging to very easy. The first question is “How challenging was it to identify policy gaps in your normal work” and participants mostly answer very challenging to somewhat challenging, with a few people answering somewhat easy. The second question asks “How challenging was it to identify policy gaps with the system” and here responses have shifted up, with a few responses in "challenging" and many more responses in the middle or at “somewhat easy”. The third question asks “How challenging was it to author policies in your normal work” and the responses here are heavily “Very challenging”. The fourth question asks “How challenging was it to author policies with the system” and here the responses are shifted up, with just a few people answering “challenging” or “somewhat challenging” and many more people answering “somewhat easy,” “easy,” or “very easy.”}
\end{figure*}

\begin{figure*}[!tb]
  \centering
  \includegraphics[width=\linewidth]{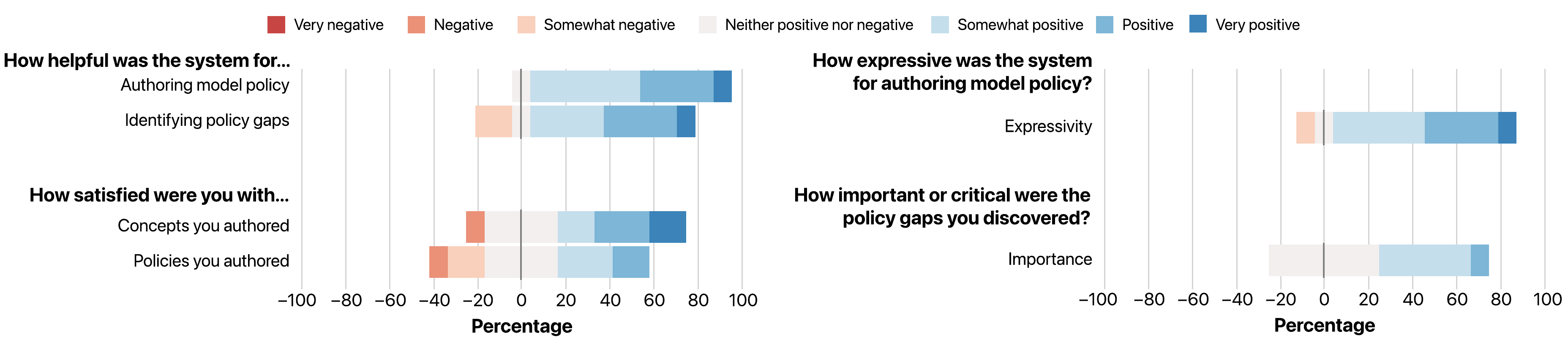}
  \caption{ 
      Participants overall found the system helpful for both identifying policy gaps and authoring model policy. They felt the system was expressive for policy authoring and tended to be satisfied with the concepts and policies they authored.
  }
  \label{fig:survey_pt2}
  \Description{Six survey questions with responses from participants. The first question asks “How helpful was the system for authoring model policy”. The second question asks “How helpful was the system for identifying policy gaps”. Participants rated authoring model policy as more helpful than identifying policy gaps. The third and fourth questions ask about how satisfied participants were with the concepts and policies they authored. Participants tended to be more satisfied with the concepts they authored than the policies they authored, and there is a somewhat smaller number of participants who felt unsatisfied or neutral with both. The fifth question asks how expressive the system was for authoring model policy. Participants mostly answered positively, with just a few answering “somewhat unexpressive” or neutral. The sixth question asks how important or critical the policy gaps you discovered were. Half the participants answer neutral on this task, with the other half answering somewhat important or important.}
\end{figure*}

\subsection{Concept \& Policy Results}
Most participants considered identifying policy gaps to be challenging in the context of their normal work (\autoref{fig:survey_pt1}). Participants started the session by brainstorming around an existing safety taxonomy, and the majority of participants (8/12) drew on specific anecdotes from their work that had previously revealed policy gaps. 
Some of these participants (4/12) also raised high-level gaps in the design of the taxonomy, such as overlap between categories or ambiguity over longer-term impacts. 
Conversely, other participants (4/12) found the taxonomy comprehensive and did not identify any policy gaps. 
With \systemName{}, all participants were successfully able to identify potential policy gaps, which extended beyond those they had expressed using the taxonomy and prior experience alone.
All participants found it easier to identify policy gaps with the system---though still sometimes challenging (\autoref{fig:survey_pt1}). 

\begin{figure*}[!tb]
  \centering
  \includegraphics[width=\linewidth]{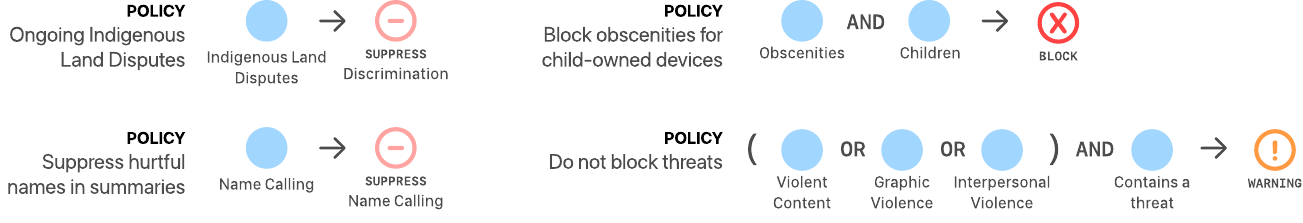}
  \caption{ 
      Using \systemName{}, participants authored a diverse set of policies that addressed a range of potentially problematic model behaviors. A few are highlighted here, with full details and all participant policies enumerated in \autoref{tab:policyListTable}. 
  }
  \label{fig:all_concepts_policies_sample}
  \Description{Four sample policies, titled “Ongoing Indigenous Land Disputes,” “Suppress hurtful names in summaries”, “Block obscenities on child-owned devices,” and “Do not block threats.”}
\end{figure*}

Participants authored a total of 24 new policies that drew upon 43 concepts, which included 12 provided safety taxonomy categories and 31 self-defined concepts (\autoref{fig:all_concepts_policies_sample}, full results in \autoref{tab:conceptListTable} and  \ref{tab:policyListTable}). Policies addressed both known and previously unknown policy gaps. 
Importantly, we had instructed participants to create desired policies from \textit{their own personal perspective} to prevent participants from limiting their creativity or restricting themselves to the set of policies they author in their professional work.
All 12 participants authored unique custom concepts that no other participant identified, and 28 of 31 concepts were distinct ideas.\footnote{The three overlapping concepts were: ``bullying,'' ``threats,'' and ``public figures,'' but we note that even with these same concept ideas, participants defined them in different ways. For example, P10 described bullying as ``Abusive, hateful content directed towards an individual with the goal of making them feel bad,'' while P12 described it as ``Content that affects the user’s mental health and state.''} Even at this constrained scale, our results reflect the value of having multiple voices engaged in the policy authoring process. We provide more detailed accounts of participants' map-exploration and map-authoring processes in \autoref{sec:map-processes}.

\subsubsection{Concepts}
Most concepts authored in the study were oriented towards policy \textit{matching conditions}. The largest set of concepts ($18/31$) described different \textit{harm categories}, such as ``animal cruelty''~(P4), ``racial slurs''~(P5), or ``cyber-bullying''~(P12). Another class of concepts ($11/31$) described \textit{high-profile or sensitive topics} where additional caution may be needed, including ``gun rights debate''~(P1) and ``general medical advice''~(P5).  Meanwhile, some participants oriented concept ideas towards policy \textit{actions}, such as sharing a crisis hotline~(P3), maintaining neutrality~(P6), or preserving the original sentiment~(P8) in output summaries.
A few practitioners used concepts to capture policy-critical context that goes beyond the case metadata available in our study, such as when the user is a child~(P3), or whether the case occurs in China~(P3) or Canada~(P10). These factors would allow important fine-grained control to enable child safety controls and policy internationalization for different locales. The full list of participant-authored concepts is provided in \autoref{tab:conceptListTable}.

\subsubsection{Policies}
As with identifying policy gaps, most participants found the task of authoring  policies to be challenging or very challenging in their normal work, but they found it easier to author policies using \systemName{} (RQ2) (\autoref{fig:survey_pt1}). A strong majority of participants found the system helpful (11/12) and expressive (10/12) for authoring model policy (\autoref{fig:survey_pt2}).
Custom concepts appeared to play an important role in policy authoring, as the majority of custom concepts that participants authored during the study session ($19/31$) were incorporated into new policies. Policies typically used 1-2 concepts in their matching conditions ($M=1.7$ concepts), and most policies were tied to blocking or warning actions (block: $n=10$, warn: $n=10$, suppress: $n=4$, add: $n=0$). See \autoref{tab:policyListTable} for the full set of policies authored by participants.

Among these policies, there were several emergent patterns in the model behaviors that participants captured and the classes of actions they chose to invoke.
First, 3/24 policies carved out scenarios that \textit{should be allowed}, but that ordinarily would be blocked (i.e., policies to avoid false positives). These include examples like ``Honor user intent while talking to partners'' (\texttt{IF: Adult sexual material AND Conversation between partners, THEN: WARN}) from P7, or ``Allow non-graphic mentions of death'' (\texttt{IF: Death, THEN: SUPPRESS Graphic violence}) from P4.
Next, 2/24 policies worked in the reverse direction to carve out scenarios that \textit{should be blocked}, but that otherwise would be allowed (i.e., policies to avoid false negatives). For example, although P3 felt that obscenities should be allowed in the general case, they authored ``Block obscenities for child-owned devices'' (\texttt{IF: Obscenities AND Children, THEN: BLOCK}) to add safeguards for children.

Another set of policies (8/24) sought to \textit{support users' well-being} by adding warnings on sensitive or risky content, but not fully blocking this content. P7 created a policy that would ``Warn for hate speech that affects mental health'' (\texttt{IF: Hate speech AND Mental health, THEN: WARN}). 
P3 created a policy ``Do not block threats'' (\texttt{IF: (Violent content AND Graphic violence AND Inter\-personal violence AND Contains a threat), THEN: WARN}) because they identified that in situations involving personal safety, the summary recipient needs to be made aware of the threat. 
A related subset of policies (4/24) similarly added blocks or warnings on sensitive topics, but with the intent of \textit{protecting the model} from controversy or potential bias. 
For example, P6 created a policy ``Ensure neutrality around the topic of Israel/Palestine`` (\texttt{IF: (Hate speech AND Discrimination AND Obscenities AND Graphic violence AND Palestine / Israel), THEN: BLOCK}), and P8 created ``Maintain sentiment from input for controversial topics'' (\texttt{IF: Controversial topics, THEN: WARN}). P2 created a policy to avoid summarizing discrimination related to ongoing indigenous land disputes (\texttt{IF: Indigenous land disputes, THEN: SUPPRESS Discrimination}). These policies mitigate risk of biased or discriminatory opinions appearing to come from the voice of the model itself rather than the text it is summarizing.
Finally, there were also more direct policies that captured clear instances of problematic content (7/24), such as ``Block sexual harassment summaries'' (\texttt{IF: Sexual Harassment, THEN: BLOCK}) from P9, ``Don't summarize disinformation'' (\texttt{IF: Disinformation, THEN: BLOCK}) from P5, and ``Block all illegal summaries'' (\texttt{IF: Illegal, THEN: BLOCK}) from P7.

\subsubsection{Expressivity}
\label{sub:expression}
Once participants started authoring policies, they ran up against limits of the grammar provided in our study interface: \texttt{AND} operators for matching conditions and \texttt{\{BLOCK, WARN, SUPPRESS, ADD\}} options for actions. 
Participants authored policies using only these operators within the short study period, but it emerged as an obvious requirement that a production-grade version of the system would need a more expressive grammar. 
The most common request for matching conditions was to support \texttt{OR} and \texttt{NOT} operators.
For example, P3 wanted to be able to author their ``Do not block threats'' policy as: (\texttt{IF: (Violent content OR Graphic violence OR Interpersonal violence) AND Contains a threat, THEN: WARN}). 
Participants also expressed interest in concepts based on external metadata, such as the type of content (e.g., email, text message) or information about the sender, recipient, and context of use.

\subsection{Participant Takeaways and Reflections}
After using \systemName{}, participants shared perspectives on how policy mapping might amplify their work and what improvements would make \systemName{} practically useful to them. 

\subsubsection*{RQ1: A map view aids understanding by bridging high-level policy with grounded examples}
Participants expressed that it was helpful to have a global view of policy~(P2, P3, P4, P5, P8, P10, P12), especially to notice potential coverage gaps~(P1, P4, P9, P10, P12). 
Concept suggestions were cited as particularly helpful to ``\textit{see our own blindspots}'' (P12) and ``\textit{identify things that we don’t have a lot of data on}'' (P1). As policy scope grows larger and harder to review, the visualization could also help to track progress towards mitigating policy gaps. For example, P10 noted that during mitigation steps, they need to answer questions about the density of training data required to address the issue, and that ``\textit{it's really cool to think of a plot like this where we could literally ensure we have the right amount of densities surrounding each policy}.''

The visual overview of the map helped participants to contextualize model failures with related examples: \textit{``not just the cherry-picked examples that people will present in policy meetings''} (P5). 
This view could serve as a boundary object across multiple stakeholders, especially those who may have differing levels of prior knowledge:
\begin{quote}
``\textit{We’re people who are deep in the trenches of policy and know it pretty clearly, but a lot of times, there’s not a single source of truth, and if you’re working with other partners, having a centralized place to capture these examples is super helpful.}'' --- P4
\end{quote}

A global view additionally helped participants to balance between micro-level issues and broader considerations (P3).
It also provided a way to ground policy ideas that are inherently abstract:
\begin{quote}
    ``\textit{The tool provides a good framework for us to visualize a lot of things that are only conceptual. I think that it becomes extremely useful because some of the policies that we author are very abstract. Here, there is more opportunity to be able to understand what some of those policies map to in terms of exact texts and data points and that is very useful to know.}'' --- P8
\end{quote}

\subsubsection*{RQ2: Custom concepts grant flexibility to address immediate policy needs}
A common reflection among participants was that they liked the ability to specify custom concepts at any granularity, and many felt this could be helpful for their work (P1, P2, P4, P7, P10, P12).
These custom concepts unlocked policies that previously would not have been possible, as P7 said: \textit{``There were times when a policy couldn't be applied to this [behavior], but I can now create a concept for it.''} P4 expressed that flexibly authoring concepts on-the-fly in \systemName{} was the biggest difference compared to their current workflow, where they would need to rely on keyword matching or expensive data labeling workflows to add new categories.

This kind of customization could be particularly helpful to support the specific needs of different AI features across the company. As P1 expressed, ``\textit{most features have their own policy, even for the same kind of content},'' and ``\textit{the most valuable part is to capture these nuances that are different}.'' Though the final policies will often differ, P10 felt that \systemName{} could provide a useful launching-off point when designing policies for new features where ``\textit{if you're building something like [Feature B], that's similar to [Feature A], and you can see [Feature A] policies and start from there.}''
Similarly, customization was very resonant with participants involved in internationalization efforts, whose work centers on adapting policies for the needs of particular regions (P2). 
Along this line, P12 expressed interest in tools like ours that could help them to differentiate among policy design decisions depending on the ``\textit{different per-locale laws and regulations and differing sensitive topics},'' which was a central challenge in their work.

\subsubsection*{RQ2: Grounded if-then policy rules can bring structure to a subjective task}
Multiple participants expressed that they liked the structure and clarity that the if-then formulation brings to the policy design task (P3, P6, P8, P10).
This contrasted against current workflows: ``\textit{The if-this, then-that, I love this. It makes the thing so objective, and these discussions are so subjective and confusing sometimes}'' (P3). 
Even for participants who had not previously thought about expressing policy via rule-like definitions, there was excitement about adopting this approach:
\begin{quote}
    ``\textit{With all the if-then conditions, that is something that could be very powerful in how we make policies to be very strong. [...] I think that is something we should really try to build on as a fundamental part of the tool.}'' --- P8
\end{quote}
While the classic if-then formulation might appear quite basic at first glance, the new expressive power lies in the ability to capture fuzzy, subtle, subjective ideas which can be directly verified by examples. Participants felt that further extensions of policy rule functionality---such as an expanded matching condition logic, an extended set of policy actions, and closer integration with their existing tools and metadata---could make the system directly useful to their work.

\subsection{Evaluation Limitations \& Future Work}
\label{eval_limitations}
This evaluation has the limitations of a preliminary study. 
Given the limited session time, some participants were initially confused about the role of concepts versus policies (P1, P10, P12). This confusion often arose if participants were building single-concept policies. Participants felt this issue could be addressed by more strongly differentiating concepts and policies on the interface.

Since we prioritized expert feedback early on in prototyping, participants used the \textit{v1} version of \systemName{} that only provided an \texttt{AND} operator to combine concepts for policy if-clauses. Due to feedback from participants that \texttt{NOT} and \texttt{OR} operations were needed (Section~\ref{sub:expression}), we added these to \textit{v2} and report all operators in Section~\ref{sec:system} to ensure future researchers know to support them all.  

Next, valuable follow-on research would integrate policy mapping within live production workflows.
Participants wondered how the tool might fit into existing collaborative policy discussions (P9, P10, P12)
and brainstormed ideas for the tool to feed into policy reviews, e.g., ``\textit{where you click submit and it proposes a new concept and or policy which could go right into our existing flows for policy reviews}'' (P10). 

Finally, experimenter bias is always a risk~\cite{barber1968fact}, and participants may have displayed a positive bias towards \systemName{} due to baseline excitement over \textit{any} new tooling support for AI policy design.
Our study design prioritized external validity to assess whether the system could aid policy designers in their work. This choice granted us a rich qualitative understanding of their policy design \textit{process}, but does not allow us to draw reliable quantitative conclusions about differences in \textit{outcomes}. Now that we have gathered promising evidence that our system can aid the policy design process, future work is needed to measure the impact of \systemName{}'s approach, which would require a counterbalanced experimental design with controlled policy design tasks.
While participants in our study represented a variety of job roles and areas of expertise, they were employed by the same organization, so they may hold shared perspectives that may not generalize to other organizations.

%% file: sections/06_results_technical_eval.tex
\label{sec:technical_eval}
\begin{figure*}[!tb]
    \centering
    \begin{subfigure}{.45\textwidth}
        \centering
        \includegraphics[width=0.7\linewidth]{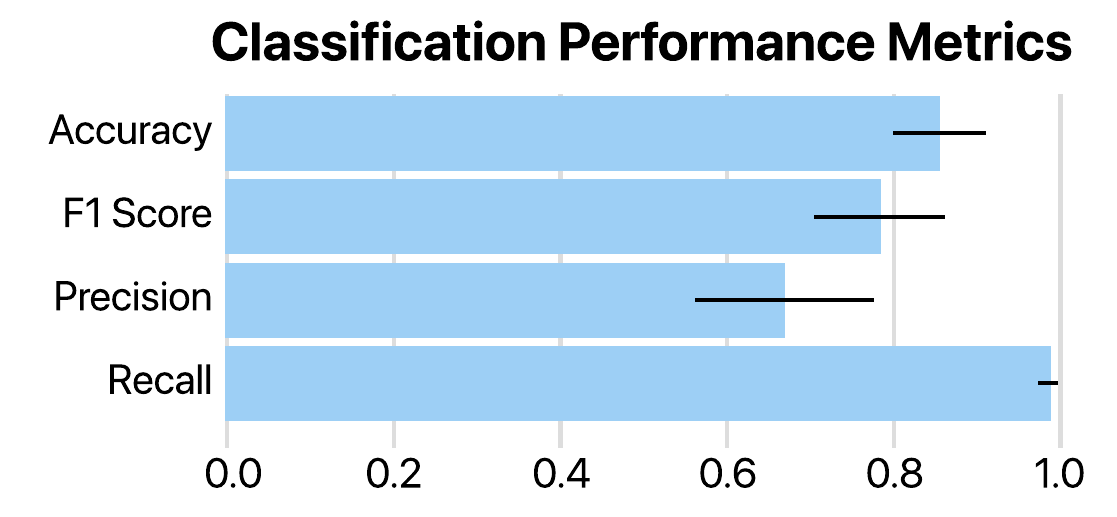}
        \caption{
            We achieve high accuracy and especially high recall across a sample of 10 participant-authored concepts \footnotesize{(bootstrapped 95\% CIs)}.
        }
        \label{fig:classif_perf}
        \Description{Horizontal bar chart with performance metrics on the y-axis and score on the x-axis. Recall is the highest score at 99\% and precision is the lowest score at 67\%.}
    \end{subfigure}%
    \hspace{1cm}
    \begin{subfigure}{.45\textwidth}
        \centering
        \includegraphics[width=0.75\linewidth]{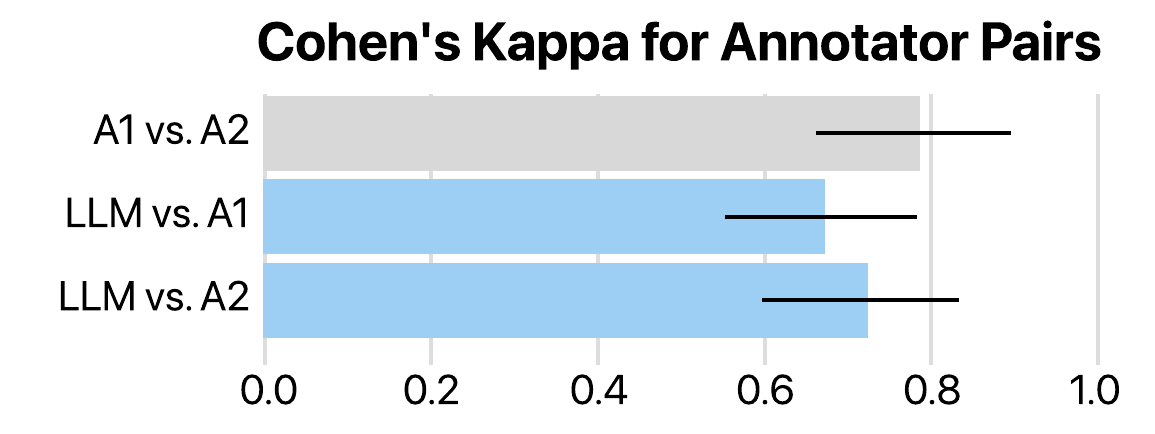}
        \caption{
            We observe comparable, moderate-to-high Cohen's $\kappa$ values between human annotators (A1 and A2) and our LLM classifier \footnotesize{(bootstrapped 95\% CIs)}.}
        \label{fig:classif_irr}
        \Description{Horizontal bar chart showing the Cohen’s kappa agreement score between two human raters, and each rater versus the LLM. The human raters have the highest level of agreement, almost 80\%, while the LLM has the lowest agreement with the first rater at 67\%.}
    \end{subfigure}
    \caption{\systemName{}'s concept classification achieves high recall of matching examples and reaches inter-rater reliability levels comparable to that of human annotators.
    }
\end{figure*}

To complement the user evaluation, we conduct technical evaluations to assess the validity of \systemName{}'s core algorithms: concept suggestion, concept classification, and model steering.

\subsection{Concept Suggestion}
Our concept suggestion method aims to surface patterns among cases that are not yet captured by existing concepts, and it builds on a previously validated toolkit~\cite{lam2024conceptInduction}. To assess the validity of suggested concepts, we use a dataset with a \textit{known} set of concepts and test whether our method can recover these ground-truth concepts. We use an Anthropic red-teaming dataset~\cite{ganguli2022redteaminglanguagemodels} that consists of LLM transcripts annotated according to a taxonomy of $n=18$ concepts such as ``Fraud \& deception,'' ``Hate speech \& offensive language,'' and ``Discrimination \& injustice''.\footnote{We exclude items in the dataset that did not match any concepts or that matched ``Other'' or ``N/A - Invalid attempt,'' which lack a distinct definition.} 
We draw a representative sample of at most 20 examples per concept to produce a dataset of $n=338$ examples (not all concepts had 20 examples). Then, we produce 5 ``partial'' versions of the concept taxonomy that randomly select 10 concepts and discard the remaining 8 concepts, which are the ground truth concepts we seek to uncover. We run our concept suggestion algorithm over the dataset labeled only with the partial taxonomy, repeating the process for three trials. Finally, we use an LLM (OpenAI's \texttt{gpt-4o-mini}) to match between the suggested concepts and ground truth concepts (prompt in \autoref{sec:appendix_prompts}).

We find that on each trial, we recover on average 40.0\% of ground truth concepts ($SD=15.1\%$), and repeated over three independent trials, we cumulatively recover on average $72.5\%$ of ground truth concepts ($SD=33.5\%$).
The method requires on average $61.4$ seconds ($SD=14.2$) and incurs a cost of \$0.13 ($SD=\$0.005$) with 173,039 tokens (input tokens: $M=163,933$; output tokens: $M=10,106$).
Among the runs, concepts that were repeatedly not recovered included ``Adult content'' and ``Conspiracy theories \& misinformation.'' Our prompt included a request for potential \textit{harms}, so these concepts may not have emerged because they are less overtly harmful compared to violence or threats.
Notably, concept suggestions that did not match known ground truth concepts may be useful additions to the existing taxonomy, such as ``Public safety threats,'' ``Harmful medical advice,'' and ``Workplace misconduct.''

\subsection{Concept and Policy Classification}
Another critical component of our system is concept classification, which underlies the if-clause of a policy. 
Our goal is to classify content in line with policy designer's intent,
so we evaluate performance for ten randomly sampled concepts authored by study participants.\footnote{Sampled concepts: ``Famous people,'' ``Severe family conditions,'' ``Death,'' ``War crimes and atrocities,'' ``Animal cruelty,'' ``Palestine / Israel,'' ``Public figures,'' ``General medical advice,'' ``Cyber-bullying,'' ``Gun rights debate.''} For each concept, we classified the full study dataset of $400$ cases using \systemName{}. This yielded an average of $12.5$ matching examples ($SD=3.9$) per concept. Next, we sampled $30$ examples per concept with up to $15$ matching examples for a total of $30*10=300$ cases. One of the authors (A1) manually annotated all 300 cases independently to gather ground truth labels. 

Across the sampled concepts, we observe a mean accuracy of $85.8\%$ ($SD=9.9\%$), with recall at $99.2\%$ ($SD=2.6\%$) and precision at $67.2\%$ ($SD=18.2\%$) (\autoref{fig:classif_perf}). Since we aim to surface potential concept candidates for user review, recall is more important than precision. Many concepts created in an AI policy context are quite subjective (e.g., users can disagree on what constitutes ``Severe family conditions''), so we compared our system's concept classifications with that of multiple human annotators. A second author (A2) independently labeled all 300 cases, and the agreement between \systemName{} and each annotator (A1 and A2) is comparable to agreement between annotators. Using Cohen's $\kappa$ as an inter-rater reliability metric, A1 vs. A2 achieve $0.79$ agreement, compared to $0.67$ and $0.73$ agreement between LLM vs. A1 and LLM vs. A2, respectively (\autoref{fig:classif_irr}).

\begin{figure*}[!tb]
    \centering
    \begin{subfigure}{.75\textwidth}
        \centering
        \includegraphics[width=\linewidth]{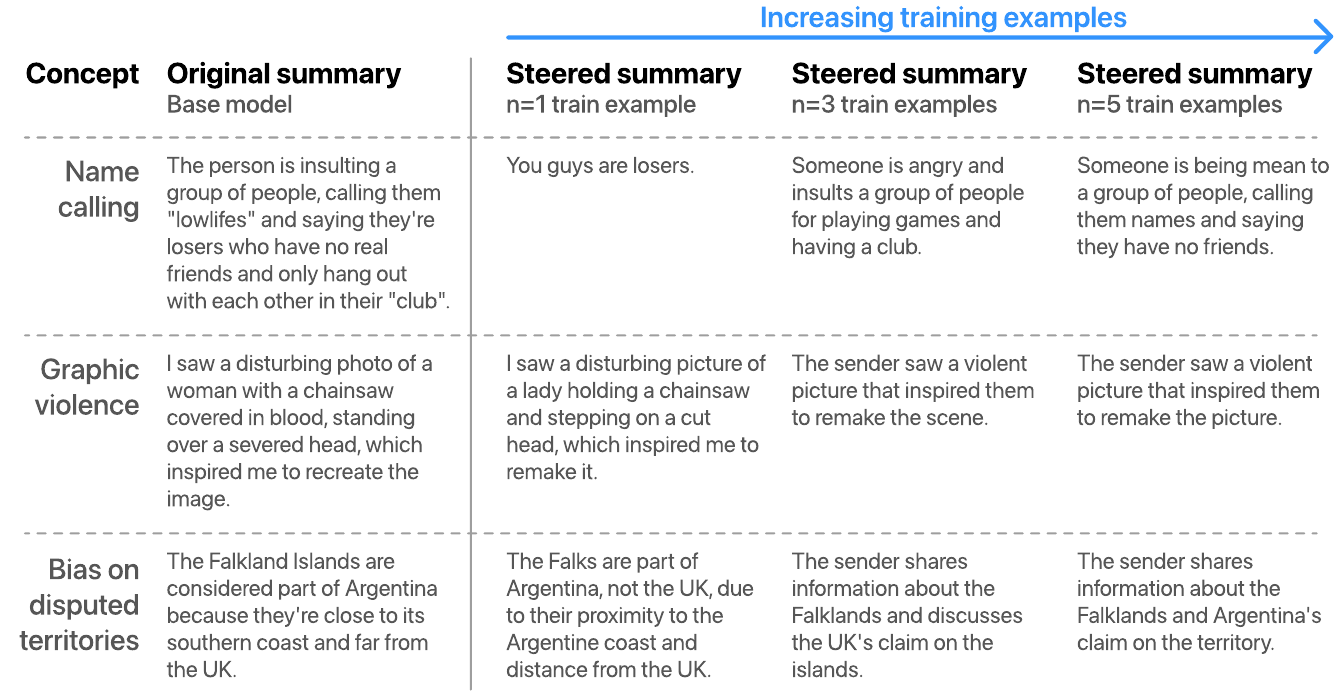}
        \caption{Compared to original summaries produced by the base model, steered summaries noticeably reduce concepts like name calling, and this reduction is more apparent with more training examples.}
        \label{fig:steering_results_text}
        \Description{A table showing examples generated by a series of models attempting to suppress a concept, across concepts “name calling”, “graphic violence”, and “bias on disputed territories.”}
    \end{subfigure}%
    \hfill
    \begin{subfigure}{.22\textwidth}
        \centering
        \includegraphics[width=\linewidth]{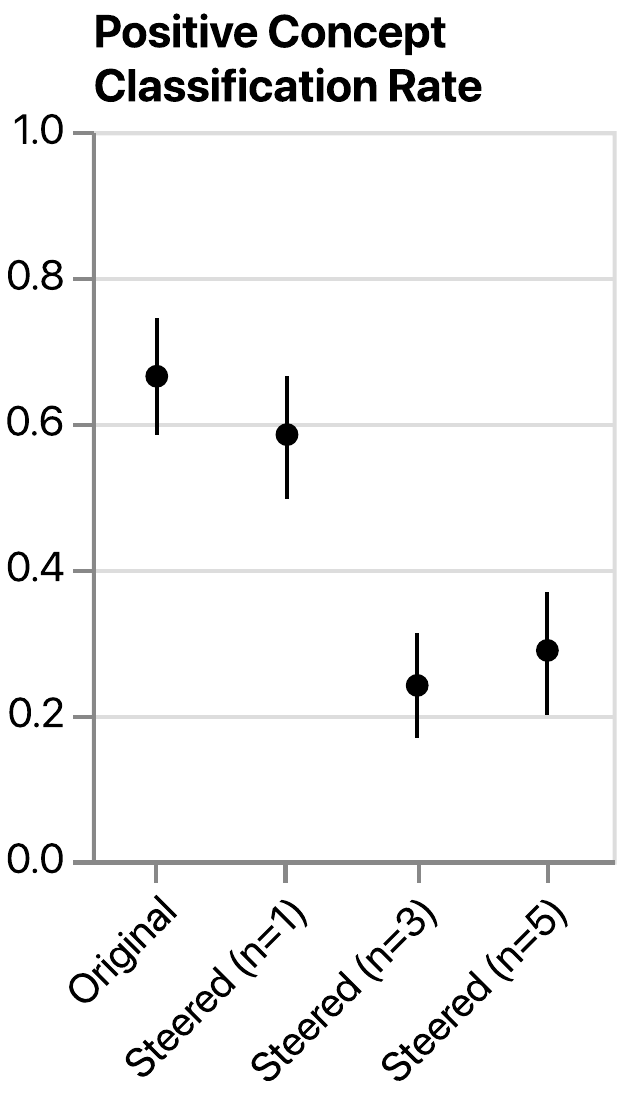}
        \caption{
        Steering significantly reduces positive concept classifications by an LLM classifier \footnotesize{(bootstrapped 95\% CIs)}.
        }
        \label{fig:steering_results}
        \Description{A plot showing the percentage of examples that are positive for a concept, against the model-steering condition. The model steered with 3 training examples performs the best, with a rate around 25\%, and the original not-steered model performs the worst, with a rate around 70\%.}
    \end{subfigure}%
    \caption{Our model steering method can successfully carry out participant-authored policy actions to suppress concepts such as ``Name calling'' and ``Graphic violence'' using as few as 3-5 training examples.}
\end{figure*}

\subsection{Model Steering}
Finally, our system provides functionality to experiment with steering the model to behave in line with a desired policy. Model steering is an active research area~\cite{wuandarora2024reft,zou2023transparency, li2024inference}, so we evaluate our approach particularly for the kinds of behaviors that participants sought for their LLM policy designs. 
First, we gather all instances of model steering policy actions from the user evaluation ($n=5$), all of which sought to suppress a specified concept: ``Name calling,'' ``Discrimination,'' ``Graphic violence,'' ``Bias on disputed territories,'' and ``Severe family conditions.'' For each concept, we gathered 10 matching examples using \systemName{}'s concept classification and split them into train and test examples. For ground truth training examples, one author (A1) manually wrote an output for each training input that shows the concept suppressed. We compare four model variants: three versions of steered models (trained with either 1, 3, or 5 examples) and the original model (as a baseline). For each model variant, we generate summaries for each test example and perform 5 trials, producing a total of 500 generations across all concepts and model variants. Finally, we use a third-party LLM (OpenAI's \texttt{gpt-4o-mini}) to independently classify whether the original concept is present in each generation, using the participant-authored concept definition. 

We find that across concepts, our steering approach results in a substantial suppression of the specified concept (\autoref{fig:steering_results}). A Wilcoxon signed-rank test\footnote{Since the input test examples were matched across model variants, for each test example, we compare concept classification results from the base model and from the steered model ($n=5$) as paired samples.} indicates that the proportion of positive concept classifications is significantly greater for the original base model ($M=0.72; SD=0.40$) than for the steered model trained with 5 examples ($M=0.31; SD=0.40$); $W = 139.0, z = -3.77, p < 0.01$.
Even with just 3 examples, we see similar levels of concept suppression in both quantitative and qualitative results (\autoref{fig:steering_results_text}).
This process is also very fast: training an intervention to suppress a specified concept requires on average 23.5 seconds ($SD=2.7$ sec), and generating a response with the steered model takes 3.5 seconds ($SD=0.2$ sec) per instruction.

%% file: sections/07_usage_scenarios.tex
\label{sec:usage_scenarios}
As a research prototype, \systemName{} was designed around a specific AI safety use case. To illustrate the broader applicability of policy maps as a concept, here we walk through several usage scenarios. 
These scenarios are fully fictional illustrations; they are not implemented in \systemName{} and do not describe an existing organization's practice. We use these usage scenarios to demonstrate how policy maps could form a foundation for a range of novel interactions beyond the current implementation.

\subsection{Multi-Stakeholder Collaboration}
Policy maps can be a useful boundary object for collaboration within LLM policy teams, across varied features and locales, and with everyday users.

\subsubsection{Live mode: Real-time deliberation}
Policy maps can augment policy discussions in real-time meetings with a \textit{live mode}. Pam has been using policy maps to design LLM policy in their daily work, but critical policy decisions are worked out in meetings with company leadership and multiple policy teams. In real-time settings, policy maps can contextualize new cases with relevant precedent.
\begin{myquote}
\scenarioSymbol{}~\textit{During a meeting, teammates discuss a new bug report involving flag-burning, so Pam pulls up live mode and enters the report as a new case. Pam sees that the case trips a current policy of warning the user on requests involving flag-burning (\texttt{IF: Flag-burning, THEN: WARN}). They also see several nearest-neighbor cases on the map that warn the user in response to requests for messages that advocate flag-burning. Pam shares these examples with the group, and a teammate points out that the new example isn't advocating for flag-burning, but is a news article on a flag-burning incident.
}
\end{myquote}

Then, policy maps can help policy designers to rapidly test out new policy ideas and discuss trade-offs.
\begin{myquote}
\scenarioSymbol{}~\textit{The group discusses a potential policy revision that only warns the user if they are advocating for flag-burning. Pam authors a new ``Advocating Flag-burning'' concept and runs the \texttt{Classify} operation on the cases that currently match the ``Flag-burning'' concept. They notice a number of cases that don't advocate for flag-burning, but are coming from countries where flag-burning is illegal. After Pam raises this area of concern, the group decides that under-flagging is riskier than over-flagging, so they keep the existing policy.
}
\end{myquote}

\subsubsection{Git for policy: Team collaboration}
Just as version control with Git allows teams of software developers to coordinate potentially-conflicting changes to large codebases, \textit{Git for policy} can support the collaborative work of policy teams. 

\begin{myquote}
\scenarioSymbol{}~\textit{Pam is investigating policy issues around food safety, so they pull the latest policy map locally. They author candidate concepts and policies to capture food safety-related harms, and once the policy looks promising, Pam drafts a pull request and assigns reviewers who have worked on related policy issues. After some back-and-forth on the review (Pam had inadvertently reverted a change to alcohol-related policies), Pam's pull request is approved and merged into the main branch.
Months later, Pam's teammate Sam is investigating a new food safety issue, and the ``blame'' history indicates that Pam last modified this policy. He enters the diff view and discovers that some underlying concept definitions have changed between the current version and Pam's version, so he makes a fix.
}
\end{myquote}

\subsubsection{Policy forks: Policy adaptation \& reuse}
Once Git for policy is set up, a policy map repository can be a useful starting point for other teams to ``fork.''

\begin{myquote}
\scenarioSymbol{}~\textit{Tammy is on a UK policy team that is about to launch new models similar to those that Pam's team develops, so she forks Pam's policy map as an initial template. She can reuse many policy rules and concepts, but she needs to alter some of concept definitions to make them appropriate for this locale (e.g., names of major political figures), and she needs to map some policy rules to a different action due to align with GDPR guidance. Months later, Pam's team pushes a major update to their policy due to a mandate from company leadership. Tammy is notified of the changes to Pam's policy map, and she ports over these critical updates.
}
\end{myquote}

\subsubsection{Participatory policy maps: External stakeholder involvement} 
Our work is targeted towards expert AI practitioners because they currently drive LLM policy efforts. However, these policies benefit from the input of diverse stakeholders~\cite{suresh2024participationInTheAge,huang2024CCAI}, which we observed even among our sample of study participants, who each contributed unique policies.
If teams are willing, they can use policy maps to support involvement from external stakeholders and everyday users.
Stakeholders can review existing policy maps to identify coverage gaps most important to them, and they can propose new concepts and policies drawing on their communities' needs.

\subsection{Model Evaluation and Auditing}
Policy maps can also aid broader evaluation and auditing efforts.

\subsubsection{Policy test suite: Longitudinal tracking}
Once a policy map has been authored, it can persist as an evaluation harness across model updates to track policy alignment.
\begin{myquote}
\scenarioSymbol{}~\textit{After working with their team to agree on a policy map, Pam adds the map to the policy test suite. They configure the system to send a report any time a new LLM checkpoint has been released. The results will also appear in an interactive dashboard where they can compare the policy map results between different model versions.
}
\end{myquote}

The policy test suite can proactively warn policy designers if a model is drifting in compliance with policies.
\begin{myquote}
\scenarioSymbol{}~\textit{A few weeks later, Pam receives an email notifying them of a regression on the policy map. Inspecting the dashboard, Pam sees substantial regressions for policies involving finance-related concepts and copyrighted entities. They raise this to the model development team and discover that the team was testing a new model compression method, which may have degraded knowledge about financial terms and copyrighted material. The team discusses a potential approach to route requests to prior model versions for these tasks with observed regressions.
}
\end{myquote}

The test suite can also notify policy designers about emerging areas of model behavior that are not covered by existing policy.
\begin{myquote}
\scenarioSymbol{}~\textit{Pam receives an email that there may be policy gaps for the most recent model, which has been deployed for several weeks. On the map, the system has highlighted two large case clusters for which there is no relevant policy. Pam finds that one cluster includes discussion about LLM jailbreaking techniques, and the other cluster involves discussion of a recent controversial Supreme Court ruling. Pam flags these as action items for their team to design new policies.
}
\end{myquote}

\subsubsection{Policy audits: Third-party model evaluation}
Finally, policy maps could be useful as an auditable artifact for external stakeholders like regulators and the public to hold model providers accountable.
An auditor can first author a policy map that captures the particular harms and biases that they wish to investigate. Then, they can gather a set of third-party models to audit and run this same policy map against all of these models' outputs. The auditor can now directly compare how well each model aligns with each policy to identify models that are particularly problematic, as well as policy issues that are especially widespread across models. The auditor can continue to run these maps over time to note policy regressions and sound the alarm when policy alignment falls below acceptable levels.

\subsection{Implementation Takeaways}
To support these scenarios, we outline several concrete implementation challenges.
While we chose LLM-based classifiers for an expressivity-speed balance, failure modes manageable at small scales become increasingly risky for for expanded datasets and timescales. Inconsistencies across runs of LLM classifiers make it challenging to reproduce prior system behavior, especially if model updates cause behavior drift. Additionally, even subtle model biases (e.g., lower performance for a dialect) could lead to major failures if all policies and concepts depend on the same LLM. One option is to use LLMs to explore new concepts, but to subsequently train smaller, traditional models for established concepts.

Collaborative and longitudinal usage scenarios surface the need for richer visualizations beyond our base map. Our system would benefit from \textit{comparative} visualizations to identify salient differences between maps, as well as \textit{temporal} visualizations to track policy drift over time. 
Notably, not all differences can be weighted equally: a minor policy rewording could be a significant issue, while a large policy rewrite could be of little importance. This means that version control, visualization, and notification systems must account for the subjective importance of policy changes, not just the presence of a change. When coordinating across many users over time, context on the significance and justifications behind policy decisions can be lost, so systems should also preserve this information.

%% file: sections/08_discussion_v1.tex
Our vision of policy mapping foregrounds the value-laden decisions inherent to LLM policy design.
Here, we discuss limitations and areas for future work.

\subsection{Reshaping LLMs with Mapmaking}
We focus on policy design in our work, but LLM evaluation and alignment processes might benefit from our mapmaking principles.

\subsubsection*{Evaluation}
For example, AI evaluation often relies on central benchmarks that embed implicit values~\cite{kiela2021dynabench, raji2021wholeWideWorldBenchmark} depending on what data samples are included and how they are labeled. 
Policy maps could be used to evaluate how well an evaluation benchmark matches our policy goals. We could even use a policy map to directly modify the samples in a dataset to better match the kinds of cases we aim to cover and our preferred labeling criteria.

\subsubsection*{Alignment}
AI alignment methods could also benefit from grounding in policy maps. Our work focuses on the tasks of reviewing existing model behavior and deciding what policies to enact. While we present methods for policy execution with blocking and steering actions, in order to support iteration at interactive speeds, we do not fully retrain or finetune model parameters. Alignment approaches such as Constitutional AI~\cite{bai2022constitutionalAI,findeis2024inverse,huang2024CCAI} and RLHF~\cite{ouyang2022training} are primarily designed for settings involving time-intensive model training. However, we could similarly experiment with using a policy map as the basis for such alignment methods by translating policies into constitutional principles, or directly adapting policy rules as automatic labelers for pairwise preference comparisons. 
Additionally, the \verb|Generate()| operator for concepts and the \verb|Act()| operator for policies could be used to generate new model training data.

\subsubsection*{Production Contexts}
We demonstrated \systemName{} to practitioners with text datasets of hundreds of cases. Ultimately, we aim to support policy design with real usage data, where we expect millions of cases across text, image, and multimodal data, so follow-on work is needed to help \systemName{} scale. Moreover, we envision that policy maps could be used directly for policy monitoring. Currently, \systemName{} does not directly measure a model's overall policy adherence. This is a limitation: the current implementation uses few-shot LLM classifiers, which are not precise to production standards and are expensive to run at scale. Using the same concept and policy abstractions, future work might substitute in different algorithms suited to large-scale settings.

\subsection{Broader Implications of Policy Maps}
We note design implications beyond the LLM domain, as well as sociotechnical implications for user participation in AI governance.

\subsubsection*{Beyond LLMs and LLM Safety}
Our work carries direct design implications for any work that grapples with normative decisions over an unbounded set of cases. 
For example, policy maps can transfer to non-LLM work in content moderation, where traditional classifiers, rule-based algorithms, and manual moderation actions are similarly based on core policies that evolve over time~\cite{fiesler2018reddit, chandrasekharan2018rules}. Policy maps might also aid human policymaking by providing a sandbox to specify and test out policy ideas before enacting them.
Beyond safety, policy maps can aid general AI system evaluation, with policies as unit tests of expected model behavior. Maps could even aid user research for product development, as concepts may serve as inspiration for emerging use cases and features.

\subsubsection*{Sociotechnical Implications}
As discussed in our usage scenarios (Section~\ref{sec:usage_scenarios}), our work presents a promising entrypoint for broader end-user participation in AI governance, as policy maps primarily require knowledge about the AI deployment domain rather than technical expertise. 
Policy maps could increase model transparency and accountability if LLM providers were compelled to publicly share their maps. While some LLM providers share high-level versions of their policies in lengthy text documents and blogs~\cite{claudeConst,openAIModelSpec}, policy maps could serve as a living document for users to readily browse and experiment with the policies that affect them most.

Policy maps also have the potential to increase stakeholder power by expanding direct involvement through map authoring. For example, LLM providers could support tailored maps based on an individual user's needs, allowing them to define their own personalized policy, in line with work exploring personalized LLMs~\cite{kirk2023personalisation, salemi2024lamp,li2024personalized}. 
We expect LLM policy to diverge for different user communities, features, and regions of the world~\cite{sorensen2024pluralisticAlignment,feng2024modularPluralism,kirk2024prism}. If users and communities could create their own policy maps, they could adapt arbitrary models towards their needs rather than relying on a centralized model platform.

%% file: sections/09_conclusion_v1.tex
LLMs elevate the challenges of AI policymaking to a new scale of complexity with an unrestricted set of potential model inputs and outputs.
We draw inspiration from the practice of mapmaking, noting that maps must also provide sound guidance without full coverage, and that they achieve this by introducing a layer of domain-specific, simplifying abstractions.
It is challenging to make simplifying assumptions in a generalized policymaking context, but we \textit{can} make decisions about which behaviors to account for or abstract away once we ground policies in specific users, tasks, and contexts of use.
We formalize this mapmaking metaphor with policy maps consisting of cases, concepts, and policy rules. These layers of abstraction allow AI practitioners to iteratively design custom maps that distill the vast realm of potential AI behavior and foreground the regions that are most critical for their particular use case. We instantiate these ideas in \systemName{}, an interactive LLM policy design tool that allows users to explore model behavior in a map visualization and update their policy map with custom concepts and policies.
Our evaluation with \participantCount{} AI safety experts demonstrates that policy maps help them to explore uncovered regions of model behavior and author grounded policies to address problematic behavior. 
Mapmaking interactions---defining custom concepts, expressing policies as structured if-then rules, and visually comparing across case, concept, and policy layers---together offer policy experts a new kind of visibility and control over the policy design process.

%% file: sections/10_appendix.tex
\section{User Evaluation Materials}
\label{sec:appendix_surveyInterviewQs}
\subsection{Post-Survey Questions}
\begin{enumerate}
    \item How \textit{helpful} was the system for identifying policy gaps? (Options: Very unhelpful, Unhelpful, Somewhat unhelpful, Neither helpful nor unhelpful, Somewhat helpful, Helpful, Very helpful)

    \item How \textit{important or critical} were the policy gaps you discovered with the system? (Options: Very unimportant, Unimportant, Somewhat unimportant, Neither important nor unimportant, Somewhat important, Important, Very important)

    \item How \textit{challenging} was it to identify policy gaps with the system? (Options: Very challenging, Challenging, Somewhat challenging, Neither easy nor challenging, Somewhat easy, Easy, Very easy)

    \item In the context of your normal work, how \textit{challenging} is it to identify policy gaps? (Options: Very challenging, Challenging, Somewhat challenging, Neither easy nor challenging, Somewhat easy, Easy, Very easy)

    \item (Optional) Were there any policy gaps you identified that you found particularly interesting or compelling? Briefly describe them here.

    \item Briefly, what did you learn or take away from this task of \textit{identifying policy gaps}?

    \item How \textit{helpful} was the system for authoring model policy? (Options: Very unhelpful, Unhelpful, Somewhat unhelpful, Neither helpful nor unhelpful, Somewhat helpful, Helpful, Very helpful)

    \item How \textit{expressive} was the system for authoring model policy? (Options: Very unexpressive, Unexpressive, Somewhat unexpressive, Neither expressive nor unexpressive, Somewhat expressive, Expressive, Very expressive)

    \item How satisfied are you with the \textit{concepts} that you authored? (Options: Very unsatisfied, Unsatisfied, Somewhat unsatisfied, Neither satisfied nor unsatisfied, Somewhat satisfied, Satisfied, Very satisfied)

    \item How satisfied are you with the \textit{policies} that you authored? (Options: Very unsatisfied, Unsatisfied, Somewhat unsatisfied, Neither satisfied nor unsatisfied, Somewhat satisfied, Satisfied, Very satisfied)

    \item How \textit{challenging} was it to author policies with the system? (Options: Very challenging, Challenging, Somewhat challenging, Neither easy nor challenging, Somewhat easy, Easy, Very easy)

    \item In the context of your normal work, how \textit{challenging} is it to author policies? (Options: Very challenging, Challenging, Somewhat challenging, Neither easy nor challenging, Somewhat easy, Easy, Very easy)

    \item (Optional) Were there any policies or concepts you authored that you found particularly interesting or compelling? Briefly describe them here.

    \item Briefly, what did you learn or take away from this task of \textit{authoring policy}?
\end{enumerate}

\subsection{Interview Questions}
\begin{enumerate}
    \item How long have you been working on responsible AI or AI safety?
    \item Does your work involve policymaking? If so, what kind of work have you done related to model policy?
    \item What is your typical workflow for policymaking? What tools or processes do you typically use to explore model behavior and author model policy?
    \item How did your experience with the system compare with your normal workflow? For example, how did your thought process, actions, or outcomes differ?
    \item Were you surprised by any of the policy gaps that were surfaced by the system?
    \item What did you like most about the system? What could be improved?
\end{enumerate}

\section{LLM Prompts}
\label{sec:appendix_prompts}

\subsection{Concept Suggestion Prompts}
\subsubsection{Distill-Summarize Prompt}
The following prompt was provided as a custom prompt for the \texttt{Distill-Summarize} LLooM operator. We provided a seed of ``harmful concepts.''

\begin{lstlisting}[language=Markdown]
I have the following TEXT EXAMPLE:
{ex}

I have this set of EXISTING CONCEPTS:
{existing_concepts}

Please summarize the aspects of this EXAMPLE that are RELATED TO {seed} and capture unique aspects of the text that are NOT captured by the EXISTING CONCEPTS. Provide the summary as at most {n_bullets} bullet points, where each bullet point is a {n_words} word phrase. Please respond ONLY with a valid JSON in the following format:
{{
    "bullets": [ "<BULLET_1>", "<BULLET_2>", ... ]
}}
\end{lstlisting}

\subsubsection{Synthesize Prompt}
The following prompt was provided as a custom prompt for the \texttt{Synthesize} LLooM operator. We provided a seed of ``harmful concepts.''

\begin{lstlisting}[language=Markdown]
I have this set of bullet point summaries of text examples:
{examples}

I have this set of EXISTING CONCEPTS:
{existing_concepts}

Please write a summary of {n_concepts_phrase} for these examples. The patterns MUST BE RELATED TO {seed}. These patterns should NOT overlap with the EXISTING CONCEPTS. For each high-level pattern, write a 2-4 word NAME for the pattern and an associated 1-sentence ChatGPT PROMPT that could take in a new text example and determine whether the relevant pattern applies. Also include 1-2 example_ids for items that BEST exemplify the pattern. Please respond ONLY with a valid JSON in the following format:
{{
    "patterns": [ 
        {{"name": "<PATTERN_NAME_1>", "prompt": "<PATTERN_PROMPT_1>", "example_ids": ["<EXAMPLE_ID_1>", "<EXAMPLE_ID_2>"]}},
        {{"name": "<PATTERN_NAME_2>", "prompt": "<PATTERN_PROMPT_2>", "example_ids": ["<EXAMPLE_ID_1>", "<EXAMPLE_ID_2>"]}},
    ]
}}
\end{lstlisting}

\subsection{Concept Classification Prompt}
The few-shot classification prompt is shown below. The zero-shot classification prompt is identical other than the exclusion of the lines for \texttt{concept\_examples}.

\begin{lstlisting}[language=Markdown]
CONTEXT:
    I have the following TEXT EXAMPLE:
    {ex}

    I have the following CRITERIA:
    {criteria}

    The following sample texts match the criteria:
    {concept_examples}

TASK:
    For the given TEXT EXAMPLE, please evaluate the CRITERIA by generating a 1-sentence RATIONALE of your thought process and providing a resulting ANSWER of ONE of the following multiple-choice options, including just the letter: 
    - A: Yes
    - B: No
    Respond with ONLY a JSON with the following format, escaping any quotes within strings with a backslash:
    {{
        "pattern_result":
            {{
                "rationale": "<rationale>",
                "answer": "<answer>",
            }}
    }}
\end{lstlisting}

\subsection{Summarization Task Prompt}
The following prompt was used with \verb|Meta-Llama-3-8B-Instruct| to generate summaries for the study dataset and model steering evaluation.

\begin{lstlisting}[language=Markdown]
Please summarize the following text into a one-sentence text message summary.
    
ORIGINAL TEXT: 
{orig}

Please only return the single one-sentence summary.
\end{lstlisting}

\subsection{Concept Suggestion Evaluation: Matching Prompt}
The following prompt was used to match suggested and ground truth concepts for the concept suggestion technical evaluation.

\begin{lstlisting}[language=Markdown]
I have this set of CONCEPTS: {ground_truth_concepts}

I have this set of TEXTS: {generated_concepts}

Please match at most ONE TEXT to each CONCEPT. To perform a match, the text must EXACTLY match the meaning of the concept.
Do NOT match the same TEXT to multiple CONCEPTS.

Here are examples of VALID matches:
- Global Diplomacy, International Relations; rationale: "The text is about diplomacy between countries"
- Statistical Data, Quantitative Evidence; rationale: "The text is about data & quantitative measures"
- Policy and Regulation, Policy; rationale: "The text is about legislation"

Here are examples of INVALID matches:
- Reputation Impact, Immigration
- Environment, Politics and Law
- Interdisciplinary Politics, Economy

If there are no valid matches, please EXCLUDE the concept from the list.
Please provide a 1-sentence RATIONALE for your decision for any matches.
Please respond with a list of each concept and either the item it matches or NONE if no item matches in this format: 
{{
    "concept_matches": [
        {{
            "concept_id": "<concept_id_number>",
            "item_id": "<item_id_number or NONE>",
            "rationale": "<rationale for match>"
        }}
    ]
}}
\end{lstlisting}

\section{\systemName{} Processes}
\label{sec:map-processes}
In addition to analyzing the final concepts and policies produced with \systemName{}, we document AI policy experts' \textit{processes} for using the system to both explore and author policy maps.

\subsection{Map Exploration Processes}
Participants used varied approaches to uncover policy gaps and concepts, and each exploration mode benefited from different features of our map visualization.
One exploration mode focused heavily on the \textit{embedding map} to perform \textit{visual checks at the global level} (P4, P6). 
For example, P4 paid attention to the placement of concepts on the map and noted concepts that were semantically similar, but were separated in the map: the conspiracy theories concept was close to controversial topics, but far away from disinformation. They felt this was a useful signal on discrepancies among concept definitions. 
Meanwhile, P6 focused on identifying outlier clusters that were distant from the existing concept markers to surface ideas for new concepts. They also performed visual checks on cluster density to estimate whether concepts were well-defined or ambiguous. For instance, they confirmed that data points for controversial topics, which they had earlier expressed were not clearly defined in the taxonomy, were scattered widely while those in illegal activity, fraud, and inauthentic practices were clustered densely. 

These explorations tended to focus on high-level patterns like concept comparison (making sense of data points within and between concepts) or concept coverage (making sense of concept outliers). In response, participants turned to suggested concepts or custom concepts to populate these regions (e.g., adding a concept for sexual harassment, which fell between existing categories of interpersonal violence and hate speech; or adding a concept for conspiracy theories, which was an outlier of controversial topics).

An alternative exploration mode centered on the \textit{tabular view} for \textit{local inspection of model behavior} (P4, P7, P8). The table view helped participants to read many examples within a concept to understand typical behavior and identify interesting trends or outliers. For example, P7 noticed concepts with inconsistent patterns of model refusals, so they reviewed examples in the table for potential patterns among input texts. 
Concepts that arose from table-centric explorations were usually inspired by specific instances of data with unique characteristics, such as explicit content between consensual partners or instances of medical advice sharing.

\subsection{Map Authoring Processes}
During the second phase of the study session, participants had the opportunity to freely author concepts or policies. We again observed that different AI safety experts favored different parts of the authoring workflow, which led to distinct authoring processes.

Some participants were most excited about the ability to author \textit{custom concepts} (P4, P7, P8). Their process tended to focus on iterative loops of (1)~exploring the data to gather new concept ideas and (2)~authoring concepts accordingly. This group of AI experts found the concept creation loop valuable because they had not previously had the flexibility to define precise model behaviors on the fly. They tended to focus on characterizing emergent model behaviors with concepts first before proceeding to policy authoring, at which point they needed to reflect on the appropriate policy specification. Perhaps because of this bottom-up focus, these participants often spent more time \textit{verifying} the behavior of their concepts and policies. For example, they ran the ``find similar cases'' and ``find policy matches'' operations so that they could carefully review and select examples that matched their intended concept or policy.

Meanwhile, some participants expressed that the \textit{if-then policy rules} were the most valuable part of the system (P3, P5, P6). These participants were often more interested in the ultimate policy outcomes they wanted to achieve, so they would often describe a policy in full before going back to author any concepts that might be needed to instantiate that policy. Given their focus on specifying a more fully-formed policy idea, these participants tended to spend less time testing their concepts or policies against data to verify their behavior. They were more interested in faithfully transferring their high-level policy ideas to a concrete instantiation in our tool, so they spent more time exploring how they might implement a policy with different possible concepts and actions.

\clearpage
\onecolumn 

\section{Concepts Created by Participants (N=31)}
 \label{tab:conceptListTable}
\begin{center}
\footnotesize
\begin{longtable}{@{}p{0.2\linewidth}  p{0.7\linewidth}@{}}
    \textbf{Concept} & \textbf{Definition}\\
    \toprule
    Mental Health & Actual planning for self harm action but not ideation\\
     \hline
    Gun rights debate & Discuss for rights to bear arms for citizens at large\\
     \hline
    Disrespect of royal monarchy & Disrespect of British royal monarch e.g. Queen Elizabeth or King Charles\\
     \hline
    Controversial Topics & Popular topics which trigger significant emotional and/or psychological responses, polarisation among large groups of people based on their ideological, geographical and cultural associations. The primary example of controversial topics are those relating to disputed territories. \\
    \hline
    Indigenous Land Disputes & Current or past land disputes on indigenous territories around the globe \\
    \hline
    Interpersonal communication harms & Model transformation that may shape human communication in ways they not intended by the originator. Unintended consequences of transforming language on communication, e.g. message, e-mails.\\
    \hline
    Display of a hotline for calls for help & Displays the national hotline phone number for the Crisis LifeLine \\
    \hline
    Modifier: Children & Allow changes on policy for children-owned device \\
    \hline
    Modifier: Country (China) & Change on policy specific to China \\
    \hline
    Modifier: contains a threat & Applies to examples where the sender is threatening the recipient \\
    \hline
    Animal cruelty & Animal cruelty, intended harm of any sentient being \\
    \hline
    Death & Any mention of death, killing, passing away, or attempted death \\
    \hline
    Public figures & Mentions or descriptions of public figures including politicians, celebrities, athletes, or fictional characters \\
    \hline
    General medical advice & Any offer of medical advice, even if it is not considered harmful or disinformation \\
    \hline
    Racial slurs & Slurs directed towards people of particular races\\
    \hline
    Palestine / Israel Controversial Topic & Content related to or mentioning the ongoing war between Palestine and Israel \\
    \hline
    Conversation between partners & Media and material that involves or includes descriptions of sexual acts, sexual references, pornography, erotica, and similar intended to arouse or stimulate sexual excitement or that promotes sexual services between consensual partner\\
    \hline
    Umbrella concept for illegal & Topics related to engaging in unlawful actions according to local, state, or federal, and other applicable laws. Promotion, selling, trafficking, or facilitation of restricted and prohibited material goods and services, and/or designed to defraud others by misrepresenting goods or services. Engaging in or enabling the sale, enslavement, or coercion of people into laborious, dangerous, or illegal actions. Attempts to use code generation capabilities to create illegal, fraudulent, or unethical outcomes.\\
    \hline
    Stereotypes & Bias, social stereotypes that are inferred during generation\\
    \hline
    Misgendering & Assumptions of gender in output that is not grounded in the output\\
    \hline
    Disability slurs & Contains a disability related slur\\
    \hline
    Name Calling & Belittling or hurtful names applied to other people to make them feel shame or embarrassment. May not be as far as hate speech and slurs.\\
    \hline
    War Crimes and Atrocities & Acts of violence committed by a warring state that are against international laws.\\
    \hline
    Sexual Harassment & Does the text mention any instance of sexual harassment or assault?\\
    \hline
    Coercion and Threats & Does the text involve coercion or threats, particularly related to intimate matters?\\
    \hline
    Canadian Controversial Topics & Controversial Topics for Canada Expansion. For example, preventing bias and hate language against those not speaking primary language of a country (Canadian English versus Canadian French). We want to prevent outputs like "Everyone in Canada should just speak English!"\\
    \hline
    Bullying & Abusive, hateful content directed towards an individual with the goal of making them feel bad.\\
    \hline
    Famous People & Does the text contain a specific name of a famous person in a given country. Is speaking about this person's activities, acts, speeches dangerous?\\
    \hline
    Severe Family Conditions & Does the text describe/shows severe or slightly obfuscated toxic conditions in a family nucleus.\\
    \hline
    Cyber-bullying & Content that affects the user's mental health and state\\
    \hline
    Revenge porn & User is being blackmailed and being threatened to have pictures of sexual nature being distributed 
\end{longtable}
\end{center}

\section{Policies Created by Participants (N=24)}
\label{tab:policyListTable}
\begin{center}
\footnotesize
\begin{longtable}{@{}p{0.2\linewidth}  p{0.4\linewidth}  p{0.3\linewidth}@{}}
    \textbf{Policy} & \textbf{Description} & \textbf{Rule}\\
    \toprule
    Block content with actual plan for self harm & Block content with actual plan for self-harm & \verb|IF| Mental Health \verb|THEN BLOCK|\\
    \hline
    Disrespect of UK monarchy & Disrespect of UK monarchy & \verb|IF| Disrespect of royal monarchy \verb|THEN WARNING| \\
    \hline
    Ongoing Indigenous Land Disputes & For land disputes on indigenous grounds around the globe, our features are not expected to respond to requests which suggest sovereignty over the lands. Sovereignty can be implied in image depiction requests or text feature requests that imply ownership by either party & \verb|IF| Indigenous Land Disputes \verb|THEN SUPRESS| Discrimination\\
    \hline
    Opt-in and out of communication language transformations & Allow user to toggle feature on and off accordingly. Always provide the option of reading the original message, before the model transformations & \verb|IF| Interpersonal communication harms \verb|THEN WARNING|\\
    \hline
    Display warning and hotline & Text that includes medical descriptions and implicit and explicit description of self-harm, summarization should be blocked, and model should offer a hotline. & \verb|IF| Regulated content \verb|AND| Self harm \verb|THEN BLOCK| and \verb|WARNING|\\
    \hline
    Block Obscenities for children-owned devices & Block Obscenities for children-owned devices & \verb|IF| Obscenities \verb|AND| Modifier: Children \verb|THEN BLOCK|\\
    \hline
    Block flag desecration for China & Flag desecration is unlawful in China, therefore flag desecration should be blocked in image and text forms & \verb|IF| Regulated Content \verb|AND| Violent Content \verb|AND| Modifier: Country(China) \verb|THEN BLOCK|\\
    \hline
    Do not block threats & It is important that the recipient, for their personal safety, is made aware that someone is threatening them, therefore threats should not be blocked. & \verb|IF| (Violent Content \verb|OR| Graphic Violence) \verb|AND| Modifier: contains a threat \verb|THEN WARNING|\\
    \hline
    Allow sexual content if not graphic & This makes a case for allowing sexual content if it's described as it'd be described in a doctor's appointment. Body parts as descriptors and regulated content as medical advice or discussion. & \verb|IF| Body Parts and Functions \verb|AND| Regulated Content \verb|THEN WARNING|\\
    \hline
    Allow non graphic mentions of death & Allow mentions of death if it's not gory, graphic, or wanton killing & \verb|IF| Death \verb|THEN SUPRESS| Graphic Violence\\
    \hline
    Don't summarize disinformation & If the input text contains content that is judged to be disinformation it should not be summarized. The user can see the full text for themselves so they can judge its full content for themselves. & \verb|IF| Disinformation \verb|THEN BLOCK|\\
    \hline
    Block only on racial slurs & Block summarization only if the input contains racial slurs; gender/religion/sexuality based slurs are OK & \verb|IF| Racial slurs \verb|THEN BLOCK|\\
    \hline
    Israel/Palestine & Ensure neutrality around this topic & \verb|IF| (Hate Speech Slurs \verb|OR| Discrimination \verb|OR| Obscenities \verb|OR| Graphic Violence) \verb|AND| Palestine/Israel - Controversial Topic \verb|THEN BLOCK|\\
    \hline
    honor user intent while talking to partners & If there is adult sexual material discussed as a part of conversation between consensual partners, we give a warning and allow the summary & \verb|IF| Adult Sexual Material \verb|AND| Conversation between partners \verb|THEN WARNING|\\
    \hline
    Warn for hate speech that affects mental health & (no description provided) & \verb|IF| Hate Speech Slurs \verb|AND| Mental Health \verb|WARNING|\\
    \hline
    Block all illegal summaries & Block all illegal summaries & \verb|IF| Umbrella concept for illegal \verb|THEN BLOCK|\\
    \hline
    maintain sentiment from input for controversial topics & Do not alter the opinion converted in the input for controversial topics & \verb|IF| Controversial Topics \verb|THEN WARNING|\\
    \hline
    Suppress hurtful names in summaries & Hurtful names in the input should not be repeated in output summaries & \verb|IF| Name Calling \verb|THEN SUPRESS| Name Calling\\
    \hline
    Block sexual harassment summaries & Sexual harassment input should not be summarized & \verb|IF| Sexual Harassment \verb|THEN BLOCK|\\
    \hline
    Quebec en\_FR Sensitivities & Bias towards Canadian English or anti-Canadian French content & \verb|IF| (Hate Speech Slurs \verb|OR| Discrimination) \verb|AND| Canadian Controversial Topics \verb|THEN WARNING|\\
    \hline
    Warn Bullies & Warn users writing bullying content & \verb|IF| Bullying \verb|THEN| WARNING\\
    \hline
    Warn the user when famous people are mentioned & Warn the user if famous people in a specific country are mentioned. It can be either their speech, acts, declarations. The latter can be fake and the user should be aware of this. & \verb|IF| Famous People \verb|THEN WARNING|\\
    \hline
    Severe Family Conditions & Avoid description of family misuse and discrimination & \verb|IF| Violent Content \verb|AND| Severe Family Conditions \verb|THEN SUPPRESS| Severe Family Conditions\\
    \hline
    Cyber-bullying & Content that affects the user's mental health and state & \verb|IF| Cyber-bullying \verb|THEN BLOCK|
\end{longtable}
\end{center}
